\documentclass[preprintnumbers,amsmath,amssymb,floatfix,11pt,prd,onecolumn,
superscriptaddress,nofootinbib]{revtex4-1}
\usepackage{latexsym}
\usepackage{epsfig}
\usepackage{epstopdf}
\usepackage{mathpazo}
\usepackage{amssymb}
\usepackage{mathtools}
\usepackage{amsmath,hyperref}
\usepackage{color}
\usepackage{graphicx}
\usepackage{graphics, setspace}

\newcommand{\mathsym}[1]{{}}
\newcommand{\unicode}[1]{{}}

\begin{document}

\title{Shadow Images of a Rotating Dyonic Black Hole with a Global Monopole Surrounded by Perfect Fluid}
\author{Sumarna Haroon}\email{sumarna.haroon@sns.nust.edu.pk}\affiliation{Department of Mathematics, School of Natural
	Sciences (SNS), National University of Sciences and Technology
	(NUST), H-12, Islamabad, Pakistan}
\author{Kimet Jusufi}
\email{kimet.jusufi@unite.edu.mk}
\affiliation{Physics Department, State University of Tetovo, Ilinden Street nn, 1200,
Tetovo, North Macedonia.}
\affiliation{Institute of Physics, Faculty of Natural Sciences and Mathematics, Ss. Cyril and Methodius University, Arhimedova 3, 1000 Skopje, North Macedonia.}
\author{Mubasher Jamil}
\email{mjamil@zjut.edu.cn}
\affiliation{Institute for Theoretical Physics and Cosmology, Zhejiang University of Technology, Hangzhou 310023, China}
\affiliation{Department of Mathematics, School of Natural
	Sciences (SNS), National University of Sciences and Technology
(NUST), H-12, Islamabad, Pakistan}

\begin{abstract}
In this paper we revisit and extend the prior work of Filho and Bezerra \cite{filho} to rotating dyonic global monopoles in presence of a perfect fluid. We then show that the surface topology at the event horizon, related to the metric computed, is a 2- sphere using the Gauss-Bonnet theorem. By choosing $\omega=-1/3, 0, 1/3$ we investigate the effect of dark matter, dust, radiation on the silhouette of black hole. The presence of the global monopole parameter $\gamma$ and the perfect fluid parameters $\upsilon$, also deforms the shape of black hole's shadow, which has been depicted through graphical illustrations. In the end we analyse energy emission rate of rotating dyonic global monopole surrounded by perfect fluid with respect to parameters.

\end{abstract}


\date{\today}
\maketitle

\section{Introduction}

Black holes are fascinating objects predicted to exist by Einstein's theory of general relativity. Recent astrophysical observation shows that such objects may exist at the center of almost every galaxy \cite{Broderick:2009ph,Broderick:2015tda}. By studying the light-like geodesics around black holes it is shown that photons can be absorbed by the black hole or can escape from black holes \cite{H1}.  That is to say a boundary is defined between these two categories of light-like geodesics, giving rise to a dark region known as the \emph{shadow}. Very recently, a project known as the Event Horizon Telescope (EHT) Collaboration announced  the first image concerning the detection of an event horizon of a supermassive black hole at the center of a giant elliptical galaxy M87  \cite{m87, Akiyama:2019eap}.
 That being said, the black hole shadow has recently become a hot topic among the researchers for the simple fact to best evaluate the soon-expected observational data. Historically, Synge was the first to propose the apparent shape of a spherically symmetric black hole \cite{Synge}. After that Luminet \cite{Luminet} discussed the appearance of a Schwarzschild black hole, the shadow of a Kerr black hole was studied by Bardeen \cite{Bardeen}, shadow of Kerr-Newman black holes \cite{devaries}, naked singularities with deformation parameters \cite{Hioki},  Kerr-Nut spacetimes \cite{abu}, while shadows of black holes in Chern-Simons modified gravity, Randall-Sundrum braneworlds, and Kaluza-Klein rotating black holes have been studied in \cite{Amarilla1,Amarilla2,Amarilla3}, and many other interesting studies concerning the effect of dark matter and cosmological constant on the shadow images \cite{Zhu:2019ura,Haroon:2018ryd,Konoplya:2019sns,shao,Xu,Zhou,Jusufi:2019nrn,Cunha:2019ikd}, Kerr-like wormholes as well as traversable wormholes and many others interesting studies \cite{Amir:2018pcu,Amir:2017slq,Shaikh:2018kfv,Shaikh:2018lcc,Gyulchev:2018fmd,Gyulchev:2019osj,Abdujabbarov:2016hnw,Abdujabbarov:2016efm,Amir:2018szm}. Some authors have also tried to test theories of gravity by using the observations obtained from shadow of Sgr A* \cite{Bambi,Bro,B1,M}. Note that new general approach for shadow calculation for axially symmetric black holes was developed in  Ref. \cite{Younsi:2016azx}, while the intrinsic curvature and topology of shadows in Kerr spacetime was developed in Refs. \cite{Wei:2018xks,Wei:2019pjf}.
 
Global monopoles are topological defects which may have been produced during the phase transitions in the early universe. In fact, global monopoles are just one type of topological defects. Other types of topological objects are expected to exist including domain walls and cosmic strings (e.g. \cite{Kibble}). More precisely, a global monopole is a heavy object characterized by spherically symmetry and divergent mass. Such objects which may have been formed during the phase transition of a system composed of a self-coupling triplet of scalar fields $\phi^a$  which undergoes a spontaneous breaking of global $O(3)$ gauge symmetry down to $U(1)$. 
The gravitational field of a static global monopole for the first time was found by Barriola and Vilenkin and are expected to be stable against spherical as well as polar perturbations \cite{birrola}. According to their model, global monopoles are configurations whose energy density decreases with the distance as $r^{-2}$ and whose spacetimes exhibit a solid angle deficit given by $\Delta=8\pi^2 \gamma^2$, where $\gamma$ is the scale of gauge-symmetry breaking. Gravitational lensing by rotating global monopoles has been investigated in Ref. \cite{Jusufi:2017lsl} and more recently in Ref. \cite{Ono:2018jrv}.  Among other things, global monopoles are expected to rotate and to carry magnetic charges.

In this paper we aim to study the impact of the rotating global monopole black hole surrounded by perfect fluid on the black hole shadow. In Section I, we consider the the gravitational field of a static dyonic black hole (SDBH) with a global monopole surrounded by perfect fluid. In Section II, by applying a complex coordinate transformation known as the Newman-Janis method \cite{Newman} we find the spacetimes of a rotating dyonic black hole (RDBH) with a global monopole surrounded by perfect fluid. In Section IV, we consider the null geodesics using Hamilton-Jacobi equation. In Section V, we study the impact of dark matter, dust and radiation on the shape of global monopole shadow.  In Section VII, we study the energy emission rate. Finally in Section VIII, we comment on our results. 

\section{A SDBH with a global monopole in perfect fluid }
The action, $S^{(EM)}$, for Einstein Maxwell gravity along with actions $S^{(D)}$ and $\mathcal{S}$ respectively defining presence of a global monopole and matter distribution, can be altogether written as
\begin{equation}\label{action}
S=S^{(EM)}+S^{(D)}+\mathcal{S}.
\end{equation} 
The Einstein-Maxwell action $S^{(EM)}$ is given by 
\begin{equation}
S^{(EM)}=\int \sqrt{-g}d^4x\left(\frac{\mathcal{R}}{2 \kappa}-\frac{1}{4}F_{\mu\nu}F^{\mu \nu}\right),
\end{equation}
where $\kappa=8\pi$. The quantities $g$, $\mathcal{R}$ and $F_{\mu\nu}$ are, respectively, the determinant of the metric $g_{\mu \nu}$ associated to the gravitational field, the scalar invariant and the electromagnetic tensor. Also $\mu$, $\nu=0,1,2,3$.\\
The corresponding Einstein field equations read
\begin{equation}\label{eq8}
\mathcal{R}_{\mu \nu}-\frac{1}{2}g_{\mu \nu}\mathcal{R}= 8 \pi T_{\mu \nu}.
\end{equation}
While the corresponding Maxwell equations are 
\begin{equation}\label{eq10}
\nabla_{\mu}F^{\mu \nu}=0.
\end{equation}
Here $T_{\mu\nu}$ is the total stress energy tensor which we discuss later in this section. Since we are considering a dyonic black hole, which means that it is comprised of both electric charge $Q_E$ and magnetic charge  $Q_M$, the electromagnetic potential has two non zero terms i.e. \cite{dyonic, Erbin}
\begin{equation}\label{eq4}
\bold{A}=\frac{Q_E}{r}dt-Q_M \cos \theta d\varphi.
\end{equation} 
The only non-vanishing components of the electromagnetic tensor 
\begin{equation}\label{eq11}
F_{tr}=-F_{rt}=\frac{Q_E}{r^2},\,\,\,F_{\theta \varphi}=-F_{\varphi \theta}=Q_M \sin \theta.
\end{equation}
Now the action $S^{(D)}$ corresponds to the matter having a defect-- a global monopole which is a heavy object formed in the phase transition of a system composed by a self-coupling scalar triplet field $\Phi^{\textbf{s}}$, where $s$ runs from $1$ to $3$. Thus the action in presence of a matter field $\Phi^{\textbf{s}}$ coupled to gravity that characterizes a global monopole \cite{birrola} 
\begin{equation}\label{eq1}
S^{\small (D)}=\int \sqrt{-g}d^4x\left(\frac{1}{2} g^{\mu\nu}\partial_{\mu}\Phi^{\textbf{s}} \partial_{\nu}\Phi^{\textbf{s}}-\frac{\lambda}{4}\left(\Phi^{2}-\gamma^2\right)^{2}\right),
\end{equation}
where $\Phi^2= \Phi^{\textbf{s}}\Phi^{\textbf{s}}$, while $\lambda$ is the self-interaction term and $\gamma$ is the scale of a gauge-symmetry breaking. The monopole can be described through the field configuration   
$\Phi^{\textbf{s}}=\frac{\gamma h(r) x^{\textbf{s}}}{|\bold{x}|}$, in which
$x^{\textbf{s}}=\left\lbrace r \sin\theta \, \cos\varphi,
 r \sin\theta \,\sin\varphi,r \cos\theta \,\right\rbrace$, such that $|\bold{x}|=r^{2} $, and $h(r)$ is a function of radial coordinate $r$.  \\
 The field equations for the scalar field $\Phi^\textbf{s}$ reduces to a single equation for $h(r)$ given as
 \begin{equation}\label{eq9}
 f(r) h''(r)+\left[\frac{2f(r)}{r}+\frac{1}{2f(r)} (f^2(r))' \right]h'(r)-\frac{2h(r)}{r^2}-\lambda \gamma^2 h(r) \left(h^2(r)-1\right)=0.
 \end{equation}
 
With these equations in mind, and without loss of generality we can choose a spherically symmetric metric written as follows
  \begin{equation}\label{SDGM}
  ds^2=-f(r)dt^2+\frac{dr^2}{f(r)}+r^2d\theta^2+r^2\sin^2 \theta d\varphi^2.
  \end{equation}
In our case the total stress-energy momentum reads
\begin{equation}\label{eq5}
T_{\mu \nu}=T_{\mu \nu}^{(EM)}+T_{\mu \nu}^{(D)}+\mathcal{T}_{\mu \nu}
\end{equation}
in which 
\begin{equation}\label{eq6}
 T_{\mu \nu}^{(EM)}=\frac{1}{4 \pi}\left(F_{\mu \sigma}{F_{\nu}}^{\sigma}    -\frac{1}{4}g_{\mu \nu}F_{\rho \sigma}F^{\rho \sigma}\right),
\end{equation}
\begin{equation}\label{eq7}
 T_{\mu \nu}^{(D)}=\partial_{\mu}\phi^{a}\partial_{\nu}\phi^a-\frac{1}{2}g_{\mu \nu}g^{\rho \sigma}\partial_{\rho}\phi^{a}\partial_{\sigma}\phi^a-\frac{g_{\mu \nu}\lambda}{4}\left(\phi^2-\gamma^2\right)^2,
\end{equation}
and $\mathcal{T}_{\mu \nu}$ is the energy-momentum tensor of the surrounding matter. The energy momentum-tensor of the surrounding fluid has the following components \cite{Heydarzade:2017wxu}
\begin{equation}\label{eq12}
{\mathcal{T}^{t}}_{t}={\mathcal{T}^{r}}_{r}=-\rho,
\end{equation}
and
\begin{equation}\label{eq13}
{\mathcal{T}^{\theta}}_{\theta}={\mathcal{T}^{\varphi}}_{\varphi}=\frac{1}{2}(1+3 \omega)\rho.
\end{equation}
Outside the core $h \to 1$ and the energy-momentum tensor of the monopole has the following components \cite{birrola}
\begin{equation}\label{eq16}
{T^ {(D)}}_{t}^{t} = {T^{(D)}}_{r}^{r} =- \gamma^{2}\left[\frac{h^{2}}{r^{2}}+f(r)\frac{(h^{\prime})^{2}}{2 }+\frac{\lambda \gamma^{2}}{4}(h^{2}-1)^{2}\right] \to - \frac{\gamma^{2}}{r^{2}},
\end{equation}
\begin{equation}\label{eq16}
{T^ {(D)}}_{\theta}^{\theta} ={T^ {(D)}}_{\varphi}^{\varphi}=-\gamma^{2}\left[f(r)\frac{(h^{\prime}(r))^{2}}{2}+\frac{\lambda \gamma^{2}}{4}(h^{2}(r)-1)^{2}\right] \to 0.
\end{equation}
The surrounding matter, whose action is denoted by $\mathcal{S}$ in Eq. (\ref{action}), can generally be a dust, radiation, quintessence, cosmological constant, phantom field or even any combination of them. 
Thus, the Einstein's field equations yield:
\begin{equation}\label{14}
\frac{r f'(r)+f(r)-1}{r^2}+\frac{8 \pi \gamma^{2}}{r^2}+\frac{Q_E^2}{r^4}+\frac{Q_M^2}{r^4}+8\pi \rho=0,
\end{equation}
\begin{equation}\label{15}
\frac{r f''(r)+2f'(r)}{2r}-\frac{Q_E^2}{r^4}-\frac{Q_M^2}{r^4}-4 \pi \rho (3 \omega +1)=0.
\end{equation}
Now by solving the set of differential equations (18) and (19) one obtains the
following general solution for the metric 
\begin{equation}\label{function}
f(r)=1-8 \pi \gamma^2-\frac{2M}{r}+\frac{Q_E^2}{r^2}+\frac{Q_M^2}{r^2}-\frac{\upsilon}{r^{1+3 \omega}},
\end{equation}
with the energy density in the form
\begin{equation}\label{eq17}
\rho=-\frac{3\, \omega\,\upsilon }{8 \pi r^{3(1+\omega)}}.
\end{equation}
Note that, $\upsilon$ is the perfect fluid parameter.  From the weak energy condition it follows the positivity of the energy density of the
surrounding field,  $\rho \geq 0$,  which should satisfy the following constraint $\omega \upsilon \leq 0$.
\section{A RDBH with a global monopole in perfect fluid}
We now extend the study of static global monopole solution and obtain its rotating counterpart. For this we apply Newman-Janis formalism to the metric (\ref{SDGM}) along with (\ref{function}). As a first step to this formalism, we transform Boyer-Lindquist (BL) coordinates $(t,r,\theta,\phi)$ to Eddington-Finkelstein (EF) coordinates $(u,r,\theta,\phi)$. This can be achieved by using the cordinate transformation
\begin{eqnarray}\label{eq18}
dt&=&du+\frac{dr}{1-8 \pi \gamma^2-\frac{2M}{r}+\frac{Q_E^2}{r^2}+\frac{Q_M^2}{r^2}-\frac{\upsilon}{r^{3\omega+1}}},
\end{eqnarray}
which yields line element in the form
\begin{eqnarray}\label{eq19}
ds^{2}&=&-\left(1-8 \pi \gamma^2-\frac{2M}{r}+\frac{Q_E^2}{r^2}+\frac{Q_M^2}{r^2}-\frac{\upsilon}{r^{3\omega+1}}\right)du^2-2dudr+r^2d\Omega^2,
\end{eqnarray}
where $d\Omega^2=d\theta^2+\sin^2\theta d\phi^2$. It's worth noting that compared to the previous work in \cite{filho}, we shall use the metric form (\ref{SDGM}) along with $f(r)$ given by Eq. (\ref{function}) to obtain a simple metric for the rotating black hole with a global monopole. This metric can be decomposed in terms of null tetrads as
\begin{eqnarray}
g^{\mu{\nu}}=-l^{\mu}n^{\nu}-l^{\nu}n^{\mu}+m^{\mu}\overline{m}^{\nu}+m^{\nu}\overline{m}^{\mu},
\end{eqnarray}
where the null vectors are defined as
\begin{eqnarray}
l^{\mu}&=&\delta^{\mu}_{r},\\
n^{\mu}&=& \delta^{\mu}_{u}-\frac{1}{2}\left( 1-8 \pi \gamma^2-\frac{2M}{r}+\frac{Q_E^2}{r^2}+\frac{Q_M^2}{r^2}-\frac{\upsilon}{r^{3\omega+1}}\right)\delta^{\mu}_{r},\\
m^\mu&=&\frac{1}{\sqrt{2}\,r}\left(\delta^{\mu}_{\theta}+\frac{\dot{\iota}}{\sin\theta}\delta^{\mu}_{\phi}\right),\\
\overline{m}^\mu&=&\frac{1}{\sqrt{2}\,r}\left(\delta^{\mu}_{\theta}-\frac{\dot{\iota}}{\sin\theta}\delta^{\mu}_{\phi}\right)
\end{eqnarray}
It is obvious from the notation that $\bar{m}^\mu$ is complex conjugate of $m^\mu$. These vectors further satisfy the conditions for normalization, orthogonality and isotropy as
\begin{eqnarray}
l^{\mu}l_{\mu}=n^{\mu}n_{\mu}=m^{\mu}m_{\mu}=\bar{m}^{\mu}\bar{m}_{\mu}=0,\\
l^{\mu}m_{\mu}=l^{\mu}\bar{m}_{\mu}=n^{\mu}m_{\mu}=n^{\mu}\bar{m}_{\mu}=0,\\
-l^{\mu}n_{\mu}=m^{\mu}\bar{m}_{\mu}=1.
\end{eqnarray}
Following the Newman--Janis prescription we write,
\begin{equation}
{x'}^{\mu} = x^{\mu} + ia (\delta_r^{\mu} - \delta_u^{\mu})
\cos\theta \rightarrow \\ \left\{\begin{array}{ll}
u' = u - ia\cos\theta, \\
r' = r + ia\cos\theta, \\
\theta' = \theta, \\
\phi' = \phi. \end{array}\right.
\end{equation}
in which $a$ stands for the rotation parameter. Next, let the null tetrad vectors $Z^a=(l^a,n^a,m^a,\bar{m}^a)$ undergo a
transformation given by $Z^\mu = ({\partial x^\mu}/{\partial {x^\prime}^\nu}) {Z^\prime}^\nu $, following
\begin{eqnarray}
l'^{\mu}&=&\delta^{\mu}_{r},\\
n'^{\mu}&=&\delta^{\mu}_{u}-\frac{1}{2}\mathcal{F}\delta^{\mu}_{r},\\\label{e11}
m'^{\mu}&=&\frac{1}{\sqrt{2\,\Sigma}}\left[(\delta^{\mu}_{u}-\delta^{\mu}_{r})\dot{\iota}{a}\sin\theta+\delta^{\mu}_{\theta}+\frac{\dot{\iota}}{\sin\theta}\delta^{\mu}_{\phi}\right],\\
\overline{m}'^{\mu}&=&\frac{1}{\sqrt{2\,\Sigma}}\left[(\delta^{\mu}_{u}-\delta^{\mu}_{r})\dot{\iota}{a}\sin\theta+\delta^{\mu}_{\theta}+\frac{\dot{\iota}}{\sin\theta}\delta^{\mu}_{\phi}\right],
\end{eqnarray}
where we replaced $f(r)$ to $\mathcal{F}(r,a,\theta)$ and $h(r)=r^2$ to $\Sigma(r,a,\theta)$. With the help of the above equations the contravariant components of new metric are computed as 
\begin{eqnarray}\notag
g^{uu}&=&\frac{a^{2}\sin^{2}\theta}{\Sigma},\   \ g^{u\phi}=\frac{a}{\Sigma}, \   \ g^{ur}=1-\frac{a^{2}\sin^2\theta}{\Sigma},\\\notag
g^{rr}&=&\mathcal{F}+\frac{a^{2}\sin^{2}\theta}{\Sigma},\   \ g^{r\phi}=-\frac{a}{\Sigma},\   \ g^{\theta\theta}=\frac{1}{\Sigma},\\
g^{\phi\phi}&=&\frac{1}{\Sigma\sin^2\theta}.
\end{eqnarray}
The new metric is found as follows
\begin{eqnarray}\notag
ds^2&=&-\mathcal{F} du^2-2dudr+2a\sin^2\theta\left(\mathcal{F}-1\right)du{d\phi}+2a\sin^2drd\phi+\Sigma d\theta^2\\ &+&\sin^2\theta\left[\Sigma+a^2\left(2-\mathcal{F}\right)\sin^2\theta\right]d\phi^2.
\end{eqnarray}
Using the method without a complexification introduced in Ref. \cite{A} we revert the EF coordinates  back to BL coordinates by using the following transformation
\begin{eqnarray}
	du=dt+\lambda(r)dr,\,\,\,\,d\phi=d\varphi+\chi(r)dr,
	\end{eqnarray}
	where
	\begin{eqnarray}
		{\lambda(r)}=\frac{-a^2-k(r)}{f(r)h(r)+a^2},\   \ {\chi(r)}=\frac{-a}{f(r)h(r)+a^2},\     \ k(r)=h(r)=r^2,
	\end{eqnarray}
	with
	\begin{eqnarray}
	\mathcal{F}&=&\frac{f(r)h(r)+a^2\cos^2\theta}{\left(k(r)+a^2\cos^2\theta\right)^2}\Sigma\label{8}
	\end{eqnarray}
Hence the rotating black hole solution in Boyer-Lindquist coordinates turns out to be
	\begin{eqnarray}\nonumber
	ds^2&=&-\frac{f(r)h(r)+a^2\cos^2\theta}{\left(k(r)+a^2\cos^2\theta\right)^2}\Sigma{dt}^2+2a\sin^{2}\theta\frac{f(r)h(r)-k(r)}{\left(k(r)+a^2\cos^{2}\theta\right)^2}\Sigma{dtd\varphi}+\frac{\Sigma}{f(r)h(r)+a^2}dr^2+\Sigma{d\theta^2}\\\nonumber
	&+&\Sigma\sin^2\theta\left[1+a^2\sin^2\theta\frac{2k(r)-f(r)h(r)+a^2\cos^2\theta}{\left(k(r)+a^2\cos^2\theta\right)^2}\right]d\varphi^2.
	\end{eqnarray}

Following \cite{A} using the condition $k(r)=h(r)=r^2$ one can find $\Sigma=r^2+a^2 \cos^2\theta$. Finally the metric can also be written as
\begin{eqnarray}\label{3}\notag
ds^2 & = & -\left(1-\frac{r^2(1-f(r))}{\Sigma}\right)dt^2
-2 a  \sin^2 \theta \left(\frac{r^2(1-f(r))}{\Sigma}   \right) dt d\varphi+\frac{\Sigma}{\Delta}dr^2+\Sigma d\theta^2\\
&+& \sin^2 \theta \left[\frac{(r^2+a^2)^2-a^2 \Delta \sin^2\theta}{\Sigma}\right]d\varphi^2.
\end{eqnarray}
where in order to simplify the notation we introduce the following quantities
\begin{eqnarray}
\Delta &=&r^2 f(r)+a^2=r^2 +a^2-2M r-8\pi r^2 \gamma^2 +Q_E^2+Q_M^2-\frac{\upsilon}{r^{3\omega-1}},
\end{eqnarray}
where $f(r)$ is given by Eq. (19). In this work, we consider three different cases of $\omega=-1/3$ dark matter dominant, 0 (dust dominant) and $1/3$ (radiation dominant).
For spin $a=0$, perfect fluid parameter $\upsilon=0$ and no charges, the above metric reduces to Schwarzschild black hole with global monopole \cite{Narayan}. It would certainly be interesting to generalize our solution by including the cosmological constant as well. As a particular example, we point out the Kerr-Newman-NUT black hole which is a subclass of the Plebanski-Demianski class, but our solution it seems not to be the case. The shadows of Kerr-Newman-NUT black holes with cosmological constant has been investigated by Grenzebach et al. \cite{ Grenzebach:2014fha}.
The electromagnetic field of a black hole is defined by its vector potential. As already mentioned, in case of a static black hole the vector potential is given by Eq. (\ref{eq4}). For the rotating case, the Newman-Janis method can also be applied on Eq. (\ref{eq4}) using a guage transformation such that $g_{rr}=0$ and $A_r=0$. For the detailed procedure the authors refer the readers to consider \cite{Erbin}. The vector potential computed through Newman-Janis formalism for a rotating dyonic black hole is given by \cite{Erbin}
\begin{eqnarray}
\bold{A}=\left(\frac{ r Q_E -a Q_M\cos\theta}{\Sigma}\right)dt+\left(-\frac{ra }{\Sigma}Q_E\sin^2\theta+\frac{r^2+a^2}{\Sigma}Q_M\cos\theta\right)d\varphi.
\end{eqnarray}
It has been shown in \cite{A}
that metric similar to (\ref{3}) satisfies the Einstein field equations. For the Einstein tensor $\emph{G}_{\mu\nu}$ and energy-momentum tensor $T_{\mu\nu}$, the Einstein field equations are given by $G_{\mu\nu}=R_{\mu\nu}-1/2 g_{\mu\nu}\,\mathcal{R}=8\pi T_{\mu\nu}$. For simplicity let  $f(r)=1-2F(r)/r^2$, where $F(r)=4 \pi \gamma^2 r^2 + M r- (Q_E^2+Q_M^2)/2+\upsilon \, r^{1-3 \omega}/2$, then the nonvanishing components of $\emph{G}_{\mu\nu}$ are
\begin{eqnarray}\nonumber
\emph{G}_{tt}&=&\frac{2}{\Sigma^3}\left(2 F(r)-\left((r^2+a^2)+a^2 \sin^2\theta\right)\right)\left(F(r)-r F'(r)\right)-\frac{a^2 \sin^2\theta}{\Sigma^2}F''(r),\\\nonumber
\emph{G}_{rr}&=&\frac{2}{\Sigma \Delta}\left(F(r)-r F'(r)\right),\\\label{einstein}
\emph{G}_{\theta\theta}&=&\frac{-2}{\Sigma}\left(F(r)-rF'(r)\right)-F''(r),\\\nonumber
\emph{G}_{t\varphi}&=&\frac{4a\sin^2\theta}{\Sigma^3}\left((r^2+a^2)-F(r)\right)\left(F(r)-rF'(r)\right)+\frac{a}{\Sigma^2}\left(r^2+a^2\right)\sin^2\theta F''(r),\\\nonumber
\emph{G}_{\varphi\varphi}&=&\frac{\sin^2\theta}{\Sigma^3}\left(4a^2 \sin^2\theta F(r)-(r^2+a^2)\left(2(r^2+a^2)+a^2\sin^2\theta\right) \right)\left(F(r)-r F'(r)\right)-\frac{(r^2+a^2)\sin^2\theta}{\Sigma^2}F''(r).
\end{eqnarray}
In terms of the orthogonal basis, for the metric (\ref{3}),
\begin{eqnarray}\label{basis}
{\emph{e}}^\mu_t&=&\frac{1}{\sqrt{\Sigma \Delta}}\left(r^2+a^2,0,0,a\right),\quad
\emph{e}^\mu_r=\frac{1}{\sqrt{\Sigma}}\left(0,1,0,0\right),\\\nonumber
\emph{e}^\mu_\theta&=&\frac{1}{\sqrt{\Sigma}}\left(0,0,1,0\right),\quad
\emph{e}^\mu_\varphi=\frac{1}{\sqrt{\Sigma \Delta}}\left(a \sin^2\theta,0,0,1\right).
\end{eqnarray}
and the Einstein tensor $\emph{G}_{\mu\nu}$, the energy momentum tensor is expressed as
\begin{eqnarray}\nonumber
p_t&=&\frac{1}{8\pi}\emph{e}^\mu_t\,\emph{e}^\nu_t \,\emph{G}_{\mu\nu},\quad
p_r =\frac{1}{8\pi}\emph{e}^\mu_r\,\emph{e}^\nu_r \,\emph{G}_{\mu\nu},\\\label{m1}
p_\theta &=&\frac{1}{8\pi}\emph{e}^\mu_\theta\,\emph{e}^\nu_\theta \,\emph{G}_{\mu\nu},\quad
p_\varphi =\frac{1}{8\pi}\emph{e}^\mu_\varphi\, \emph{e}^\nu_\varphi\, \emph{G}_{\mu\nu}.
\end{eqnarray}
Eqs. (\ref{3}-\ref{m1}) gives the components for energy momentum tensor as
\begin{eqnarray}
p_t&=&\frac{1}{8\pi\Sigma^2}\left(8 \pi \gamma^2 r^2 - 3\upsilon \omega r^{1-3 \omega}+(Q_E^2+Q_M^2)\right)=-p_r,\\\nonumber
p_\theta&=&-p_r-\frac{1}{8\pi\Sigma}\left(8 \pi \gamma^2 - \frac{3 \upsilon \omega (1-3 \omega)}{2 r^{1+3 \omega}}\right)=p_\varphi.\\\nonumber
\end{eqnarray}
Analogous to Kerr black hole, a ring singularity resides inside the black hole defined by metric (\ref{3}). This can be demonstrated by computing the points at which the Kretschmann scalar $\mathcal{K}_s=R_{\mu\nu\sigma\rho}R^{\mu\nu\sigma\rho}$ turns to infinity. For the metric (\ref{3}), the Kretschmann scalar has the value
\begin{equation}
\mathcal{K}_s=\frac{Z(r, a,\theta, Q_E, Q_M, \omega, \upsilon ,\gamma)}{r^{2(1+3\omega)}(r^2+a^2\cos^2\theta)^2}.
\end{equation}
where $Z(r, a, \theta, Q_E, Q_M, \omega, \upsilon, \gamma)$ is a tedious function. From the above expression, we observe that for $\omega=-1/3,0,1/3$ the poles lie $r^2+a\cos^2\theta=0$ or when $r=0$ and $\theta=\pi/2$. This leads us to the interpretation that a test particle moving in an equatorial plane $\theta=\pi/2$ will hit the singularity at $r=0$.

\subsection{Surface Topology}
It is interesting to determine the surface topology of the global monopole spacetime at the event horizon. At a fixed moment in time $t$, and a constant $r=r_+$, the metric (\ref{3}) reduces to
\begin{eqnarray}
ds^2 & = & \Sigma(r_+,\theta) d\theta^2+\left(2Mr_++8 \pi r_+^2 \gamma^2-Q_E^2-Q_M^2+\frac{\upsilon}{r_+^{3 \omega-1}}\right)^2 \frac{\sin^2\theta}{\Sigma(r_+,\theta)} d\varphi^2,
\end{eqnarray}
The above metric has the following determinant
\begin{eqnarray}
\det g^{(2)}&=& \left(2Mr_++8 \pi r_+^2 \gamma^2-Q_E^2-Q_M^2+\frac{\upsilon}{r_+^{3 \omega-1}}\right)^2 \sin^2\theta.
\end{eqnarray}

\textbf{Theorem}: \textit{Let $\mathcal{M}$ be a compact orientable surface with metric $g^{(2)}$, and let $K$ be the Gaussian curvature with respect to $g^{(2)}$ on $\mathcal{M}$. Then, the Gauss-Bonnet theorem states that }
\begin{equation}
\iint_{\mathcal{M}} K \,dA=2 \pi \chi(\mathcal{M}).
\end{equation}
Note that $dA$ is the surface line element of the 2-dimensional surface and $\chi(\mathcal{M})$ is the Euler characteristic number. It is convenient to express sometimes the above theorem in terms of the Ricci scalar, in particular for the 2-dimensional surface there is a simple relation between the Gaussian curvature and Ricci scalar given by
\begin{equation}
K=\frac{\mathcal{R}}{2}.
\end{equation}
Yielding the following from
\begin{equation}
\frac{1}{4 \pi}\iint_{\mathcal{M}} \mathcal{R} dA=\chi(\mathcal{M}).
\end{equation}

A straightforward calculation using the  metric (49) yields the following result for the Ricci scalar
\begin{equation}
\mathcal{R}=\frac{2 (r_+^2+a^2 ) (r_+^2-3a^2 \cos^2 \theta)}{\left(r_+^2+a^2 \cos^2 \theta\right)^3}
\end{equation}

From the GBT we find
\begin{equation}
\chi(\mathcal{M})=\frac{1}{4 \pi}\int_{0}^{2 \pi}\int_{0}^{\pi} \left[\frac{2 (r_+^2+a^2 ) (r_+^2-3a^2 \cos^2 \theta)}{\left(r_+^2+a^2 \cos^2 \theta\right)^3} \right] \sqrt{\det g^{(2)}} d\theta d\varphi.
\end{equation}

Finally, solving the integral we find
\begin{equation}
 \chi(\mathcal{M})=2.
\end{equation}

Hence the surface topology is a 2-sphere at the event horizon, since we know that $\chi(\mathcal{M})_{sphere}=2$.

\begin{figure*}
\includegraphics[scale=0.64]{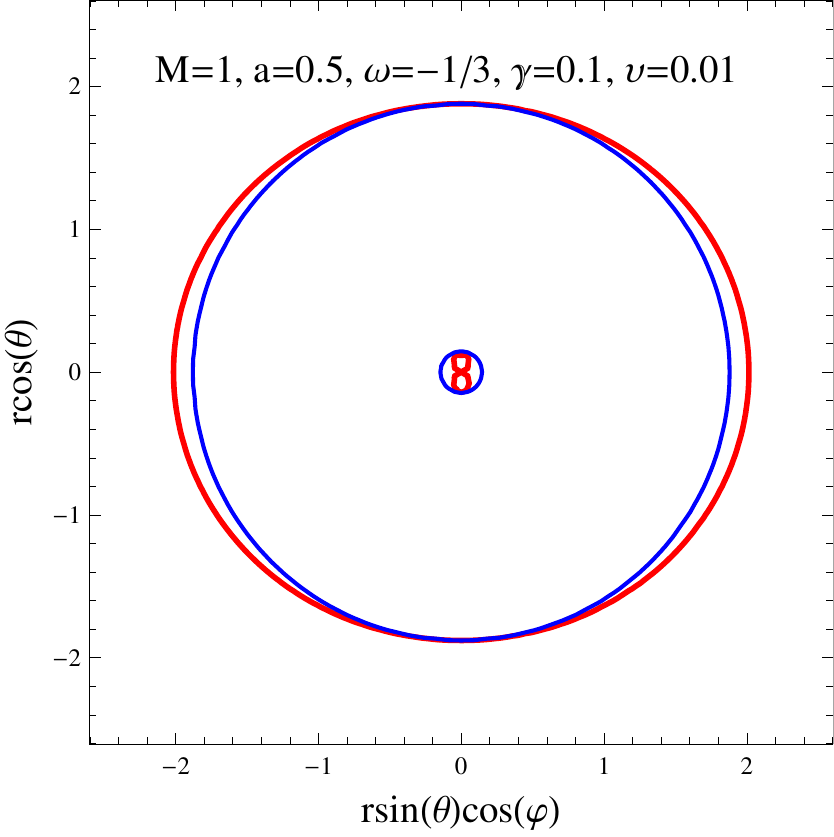}
\includegraphics[scale=0.64]{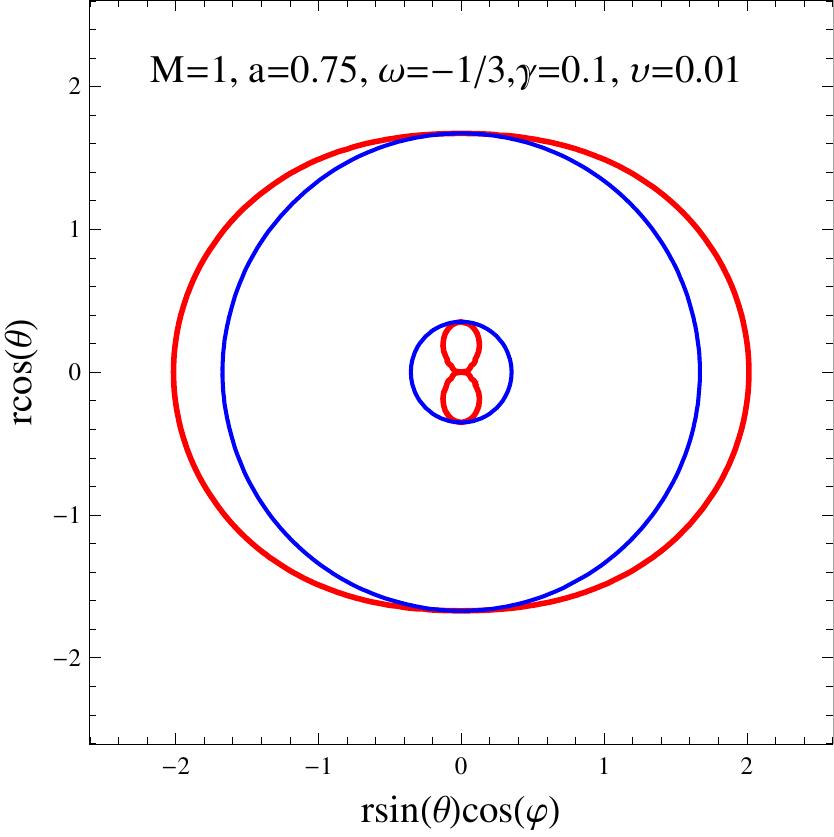}
\includegraphics[scale=0.64]{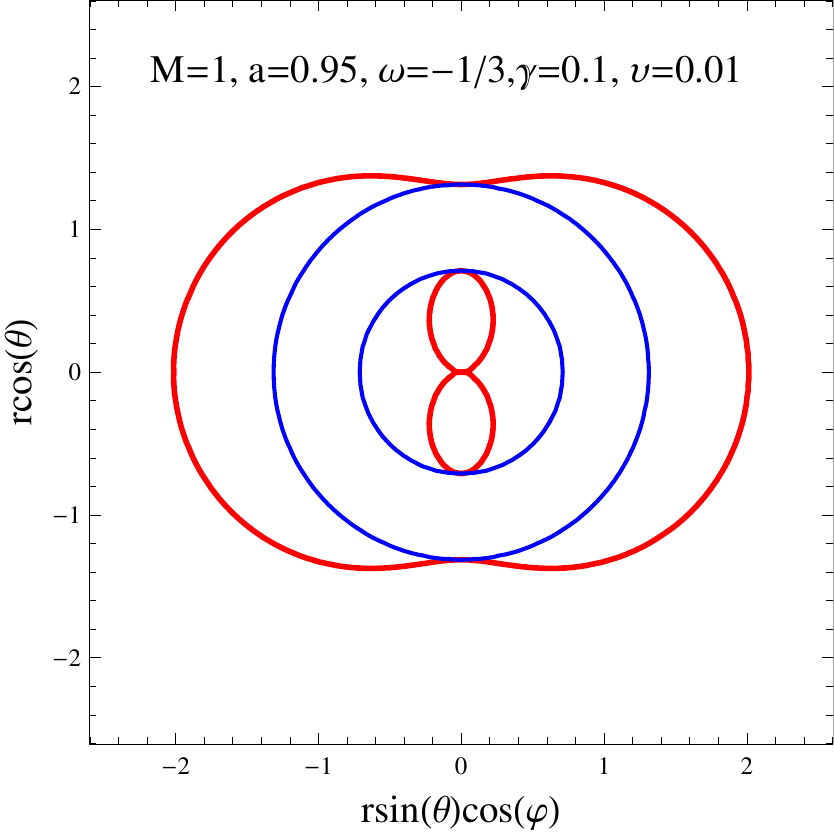}\\
\includegraphics[scale=0.64]{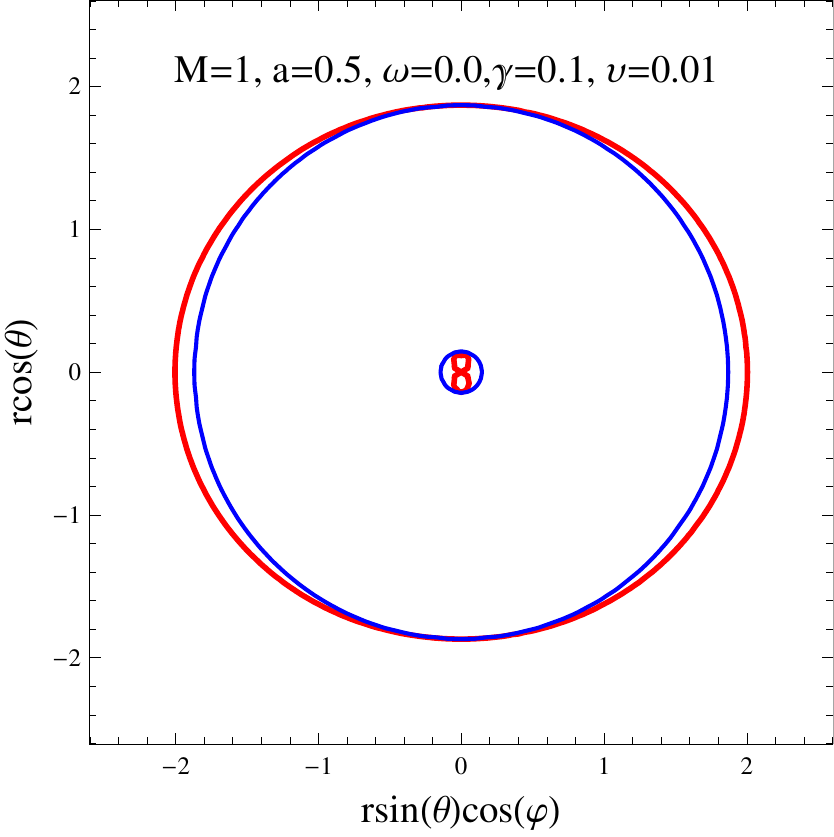}
\includegraphics[scale=0.64]{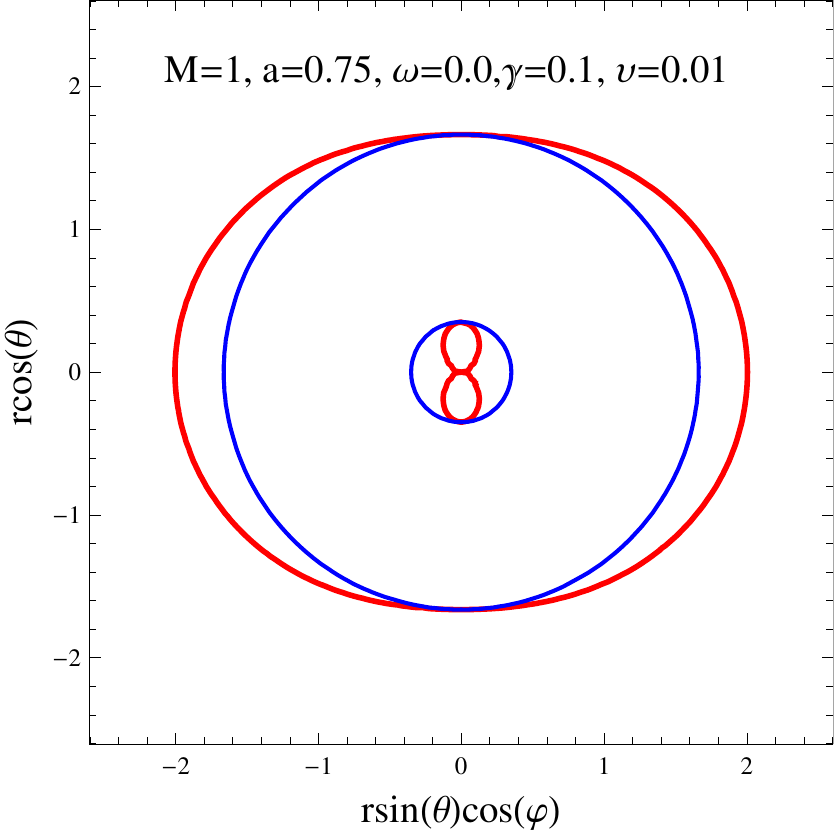}
\includegraphics[scale=0.64]{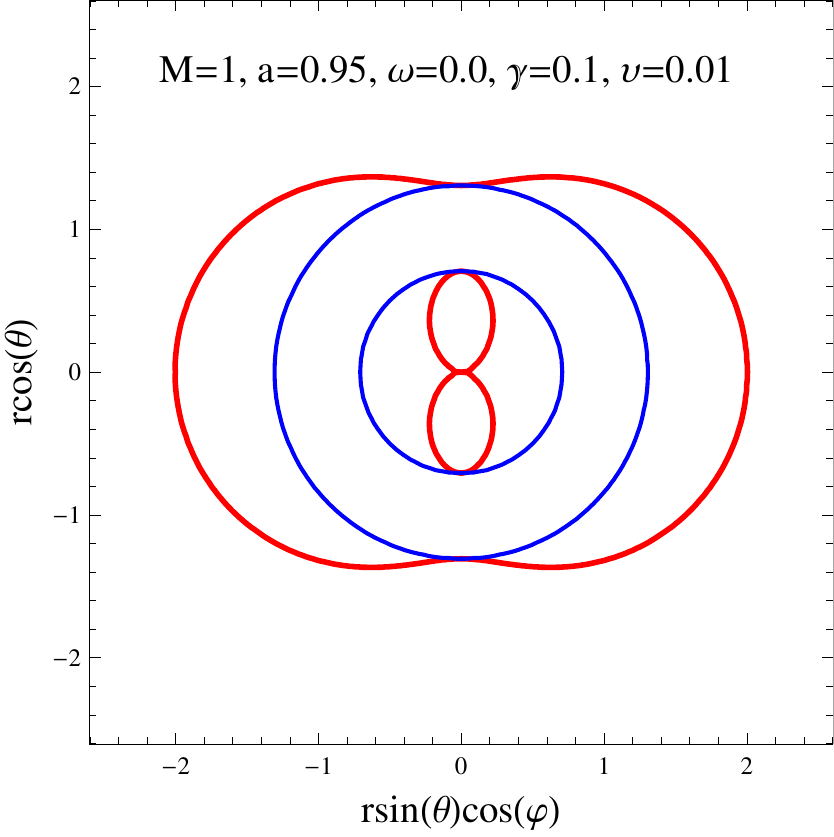}\\
\includegraphics[scale=0.64]{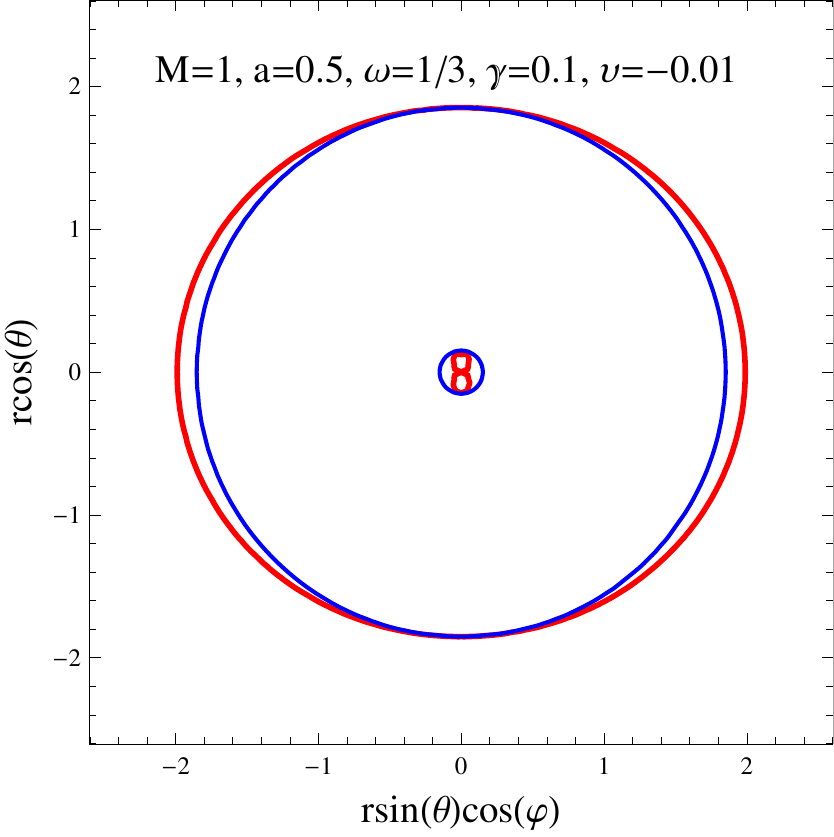}
\includegraphics[scale=0.64]{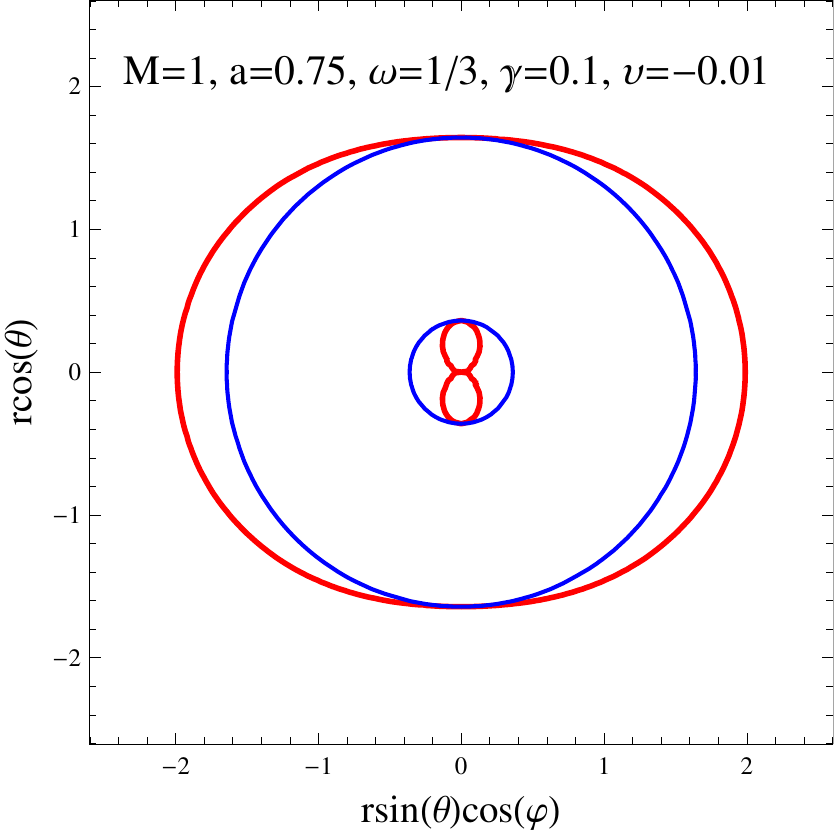}
\includegraphics[scale=0.64]{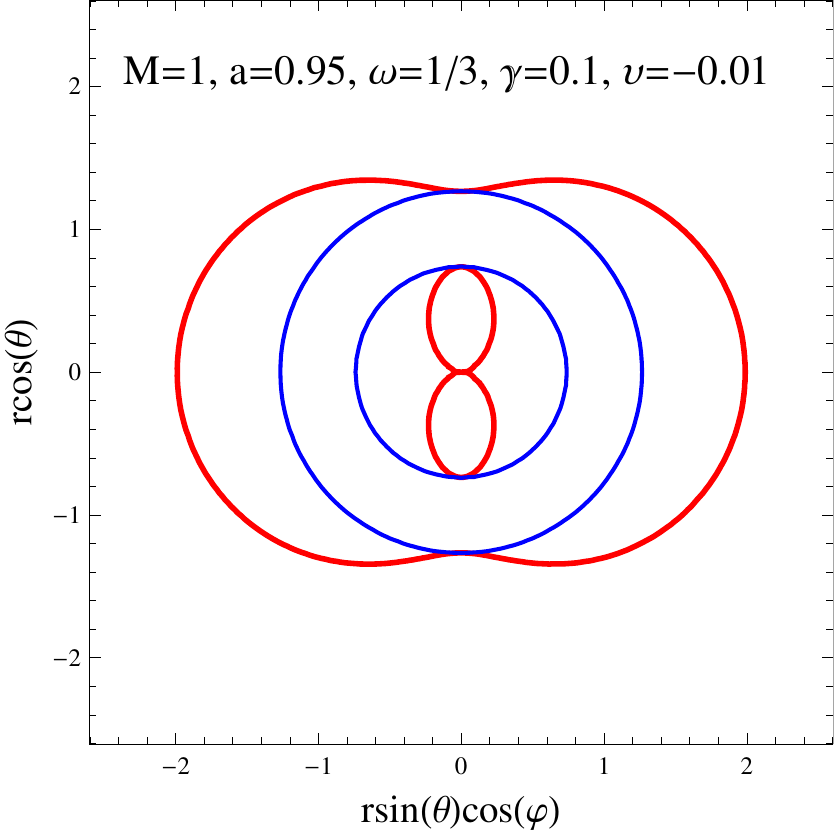}\\
\caption{\label{shad} Plots showing the shape of ergoregion in $xz$-plane for
different values of $a$, $\omega$, and $\upsilon$. We have chosen $Q_E=Q_M=0.1$ in all plots. The blue and the red lines correspond to horizons and static limit surfaces, respectively. The outer red line corresponds to the static limit surface,
whereas the two blue lines correspond to the two horizons. Due to the small values of $\upsilon$ and a arbitrary value of $a$ we observe almost indistinguishable plots for the shape of the ergoregion.}
\end{figure*}
\subsection{Shape of Ergoregion}
Let us now proceed to study the shape of the ergoregion of a RDBH with a global monopole. In particular we shall be interested to plot the shape of the ergoregion in the $xz$-plane. Recall that the horizons of the RDG can be found by solving $\Delta=0$, 
\begin{equation}
r^2+a^2-2Mr-8 \pi r^2 \gamma^2 +Q_E^2+Q_M^2-\frac{\upsilon}{r^{3 \omega-1}}=0,
\end{equation}
on the other hand, the limit surfaces or inner and outer ergosurface is given by $g_{tt}=0$, i.e., 
\begin{equation}
r^2+a^2 \cos^2\theta-2Mr-8 \pi r^2 \gamma^2 +Q_E^2+Q_M^2-\frac{\upsilon}{r^{3 \omega-1}}=0.
\end{equation}

There is an interesting process which relies on the presence of an ergoregion, namely from such a rotating black hole energy can be extracted, and is known as the Penrose process. In Figure 1 we plot the shape of ergoregion for different values of $a$, $\omega$, $\gamma$, and $\upsilon$. One can observe that the event horizon and static limit surface meet at the poles while the region between them is the ergoregion which supports negative energy orbits. Furthermore the shape of ergoregion, depends on the spin $a$, however due to the small values of $\upsilon$ we observe small changes related to the value of $\omega$.
\section{NULL GEODESICS}
Our main objective is to study the shadow cast by the black hole defined by metric (\ref{3}). To do so, we first need to analyze the geodesics structure of photons moving around the compact gravitational source. This will enable us to detect the unstable photon orbits which in turn defines the boundary of the shadow.\\
To observe the null geodesics around the RDGM present in perfect fluid, we consider the Hamilton-Jacobi method. The Hamilton-Jacobi equation is given by 
\begin{equation}
\partial_\tau\mathcal{J}=-\mathcal{H}.
\label{HJ1}
\end{equation}
In the above equation
\begin{description}
  \item[On Left Side] $\mathcal{J}$ is the Jacobi action, defined as the function of affine parameter $\tau$ and coordinates $x^\mu$ i.e. $\mathcal{J}=\mathcal{J}(\tau,x^\mu)$.
  \item[On Right Side] $\mathcal{H}$ is the Hamiltonian of test particle's motion and is equivalent to  $g^{\mu\nu}\partial_\mu\mathcal{J}\;\partial_\nu\mathcal{J}$.
\end{description}
In the spacetime under consideration, along the photon geodesics the energy $E$ and momentum $L$, defined respectively by Killing fields $\xi_t=\partial_t$ and $\xi_\phi=\partial_\phi$, are conserved. The mass $m=0$ of the photon is also constant. Using these constants of motion we can thus separate the Jacobi function as
\begin{equation}
\mathcal{J}=\frac{1}{2}m^2\tau-Et+L\phi+\mathcal{J}_r(r)+\mathcal{J}_{\theta}(\theta)
\label{HJ2}
\end{equation}
where the functions $\mathcal{J}_r(r)$ and $\mathcal{J}_\theta(\theta)$ respectively depends on coordinates $r$ and $\theta$.
 \begin{figure}[h!]
 \includegraphics[width=5.cm,height=5.cm]{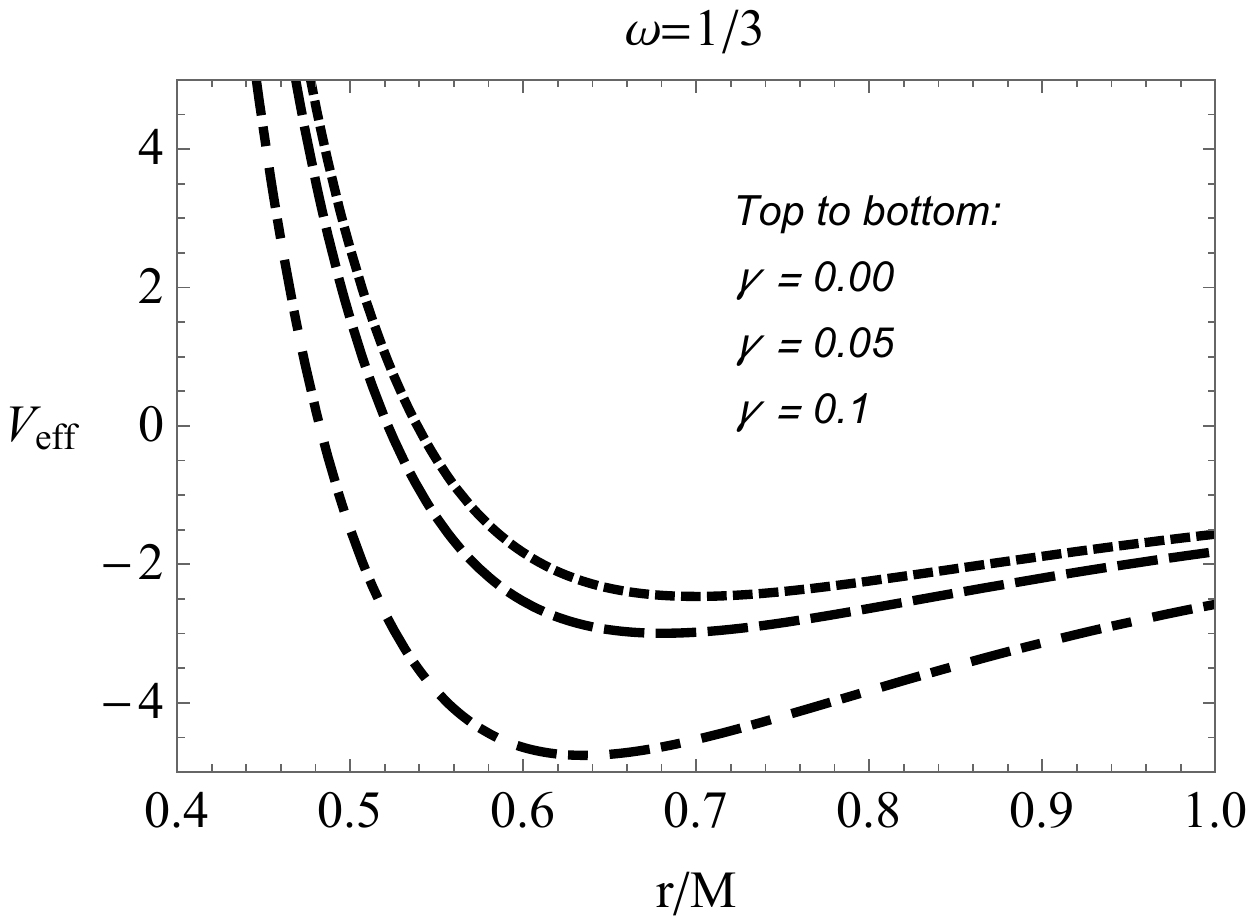}
 \includegraphics[width=5.cm,height=5.cm]{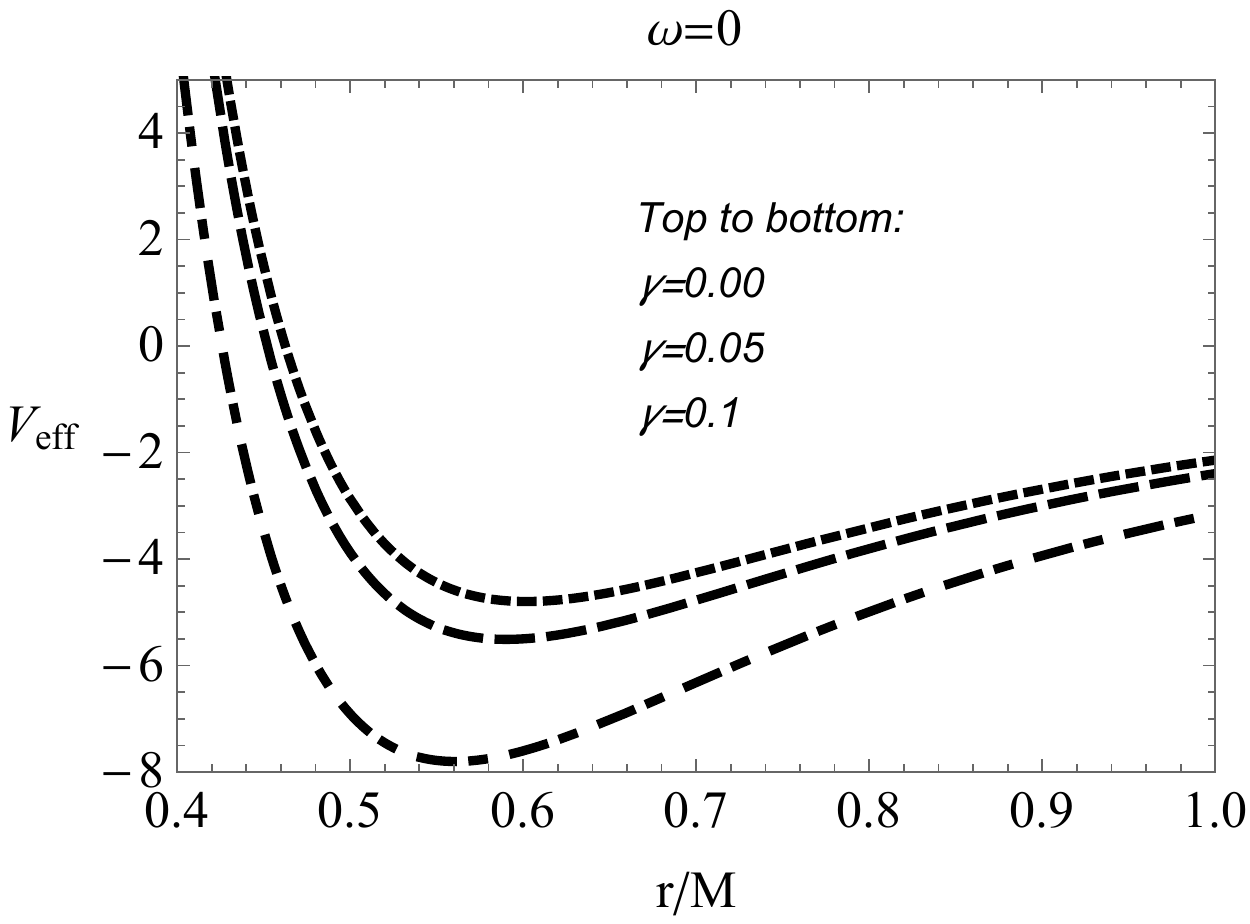}
  \includegraphics[width=5.cm,height=5.cm]{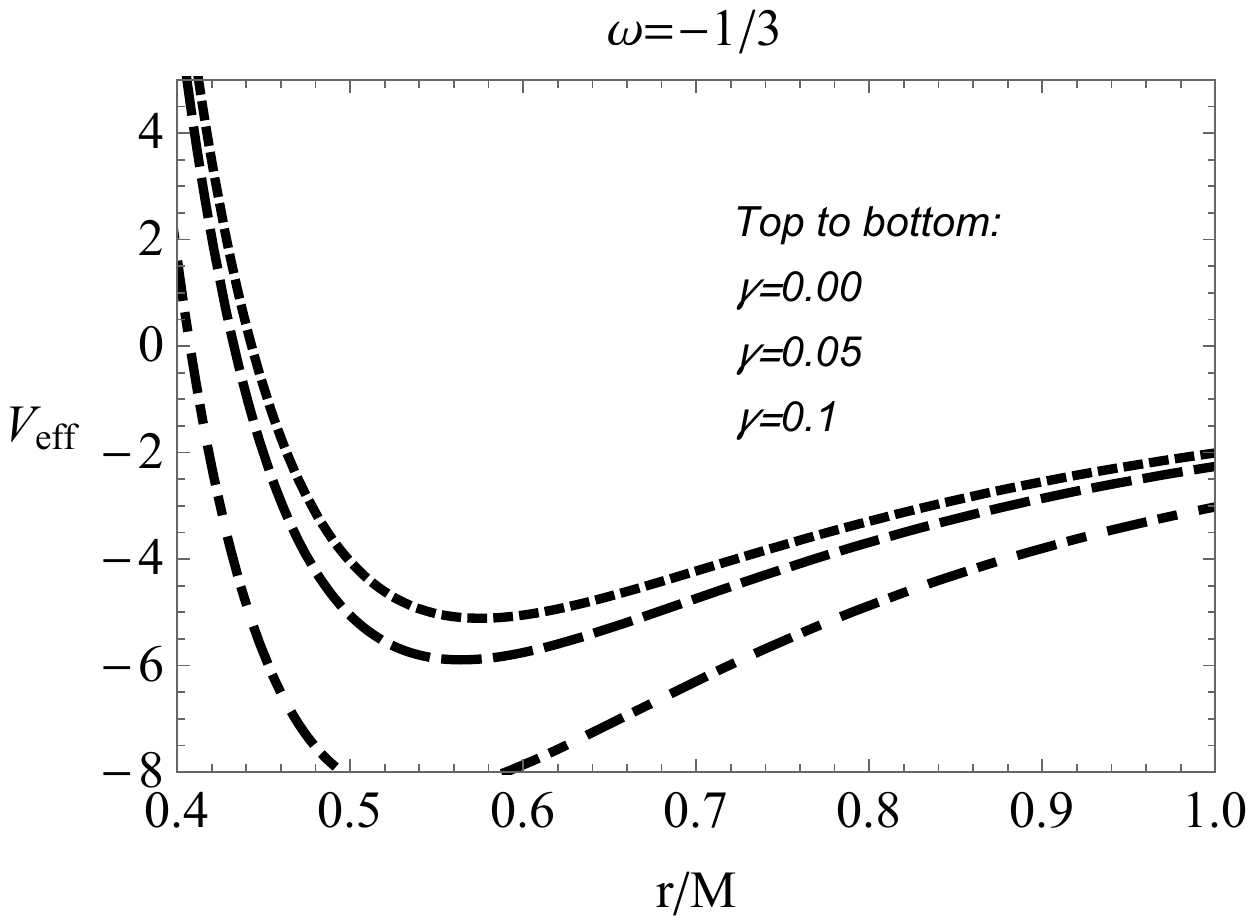}
  \caption{\label{figure0}The effective potential of photon moving in equatorial plane, with respect to its radial motion: $\omega=1/3$ for radiation, $\omega=0$ for dust and $\omega=-1/3$ for dark matter.}
  \end{figure}
Combining Eq. (\ref{HJ1}) and Eq. (\ref{HJ2}) yields a set of equations, which describes the dynamics of a test particle around the rotating black hole in perfect fluid matter, as:
\begin{align}
\label{HJ3}
&\Sigma\frac{dt}{d\tau}=\frac{r^2+a^2}{\Delta}[E(r^2+a^2)-aL]-a(aE\sin^2\theta-L),\\
&\Sigma\frac{dr}{d\tau}=\sqrt{\mathcal{R}(r)},\\
&\Sigma\frac{d\theta}{d\tau}=\sqrt{\Theta(\theta)},\\
\label{HJ4}
&\Sigma\frac{d\varphi}{d\tau}=\frac{a}{\Delta}[E(r^2+a^2)-aL]-\left(aE-\frac{L}{\sin^2\theta}\right),
\end{align}
where $\mathcal{R}(r)$ and $\Theta(\theta)$ read as
\begin{align}
\label{HJ5}
&\mathcal{R}(r)=[E(r^2+a^2)-aL]^2-\Delta[m^2r^2+(aE-L)^2+\mathcal{K}],\\
&\Theta(\theta)=\mathcal{K}-\left(  \dfrac{L^2}{\sin^2\theta}-a^2E^2  \right) \cos^2\theta,
\end{align}
with $\mathcal{K}$ the Carter constant.
\section{Circular Orbits}
Now we consider a gravitational source placed between a light emitting source and an observer at infinity. The photons emitted from the light source will form two kinds of trajectories: the ones which eventually fall into the black hole and the ones which scatter away from it. The region separating these trajectories, contains unstable circular orbits. These unstable circular orbits form a dark region in sky thus forming the contour of the shadow. In this section we intend to discuss the presence of unstable circular orbits around the black hole under consideration. For this we consider photon as a test particle and hence take $m=0$. We can express the radial geodesic equation in terms of effective potential $V_{\text{eff}}$ of photon's radial motion as
\begin{equation*}
\Sigma^2\left(\frac{dr}{d\tau}\right)^2+V_{\text{eff}}=0.
\end{equation*}
For our convenience we introduce two independent parameters $\xi$ and $\eta$ \cite{Chandrasekhar:1992} as
\begin{equation}
\xi=L/E,  \;\; \;\; \;\;   \eta=\mathcal{K}/E^2.
\end{equation}
The effective potential in terms of these two parameters is then expressed as
\begin{equation}\label{veff}
V_{\text{eff}}=\Delta((a-\xi)^2+\eta)-(r^2+a^2-a\;\xi)^2,
\end{equation}
where we have replaced $V_{\text{eff}}/E^2$ by $V_{\text{eff}}$. Figure (\ref{figure0}) shows the variation in effective potential associated with the radial motion of photons. From the figure we observe that in all three cases the value of effective potential decreases with increase in parameter $\gamma$. Now the circular photon orbits exists when at some  constant $r=r_c$ the conditions
\begin{equation}\label{cond}
V_{\text{eff}}(r)=0,\quad~~~\frac{dV_{\text{eff}}(r)}{dr}=0
\end{equation}
are satisfied. We then use Eq. (\ref{veff}) in Eq. (\ref{cond}) and thus obtain
\begin{align}
\label{R2}
&[\eta+(\xi-a)^2]\Delta-(r^2+a^2-a\xi)^2=0,\\
&4(r^2+a^2-a\xi)-[\eta+(\xi-a)^2]\mathcal{A}(r)=0,
\label{R3}
\end{align}
where
\begin{equation}\label{aa}
\mathcal{A}(r)= 1- 8\pi \gamma^2-\frac{M}{r}+(\frac{3 \omega -1}{2})\frac{\upsilon}{r^{1+3\omega}}.
\end{equation}
Combining Eqs. (\ref{R2}-\ref{R3}) results in
\begin{align}
&a\;\xi=r^2+a^2-\frac{2\;\Delta}{\mathcal{A}(r)}\;,\\
&\eta=\frac{4 \Delta}{\mathcal{A}(r)^2}-\frac{1}{a^2}\left(r^2-\frac{\Delta}{\mathcal{A}(r)^2}\right)^2
\end{align}
It is worth mentioning here that impact parameters, $\xi$ and $\eta$, will be affected not just by radial coordinate $r$, spin parameter $a$ and mass of black hole $M$ but also by electric charge $Q_E$, magnetic charge  $Q_M$, monopole parameter $\gamma$ and perfect fluid parameter $\upsilon$.
The unstable circular orbits are located at local maxima of the potential curves i.e. when $V_{\text{eff}}^{''}<0$ or
\begin{equation}
\left({\Delta^\prime}^2+2 \Delta \Delta^{\prime\prime}\right)r+ 2 \Delta \Delta^\prime>0
\end{equation}
\section{SILHOUETTE OF BLACK HOLES }
\label{shadow}
In this section, we extend our calculations to observe shadow of RDGM surrounded by perfect fluid. To gain the optical image we specify the observer at position $(r_o,\theta_o)$, where $r_o=r\rightarrow\infty$ and $\theta_o$ is the angular coordinate at infinity, on observer's sky. The new coordinates, also widely known as celestial coordinates, $\alpha$ and $\beta$ are then introduced. These coordinates are selected such that $\alpha$ and $\beta$ correspond to the apparent perpendicular distance of the image from axis of symmetry and its projection on the equatorial plane, respectively.

Due to the presence of global monopole we have asymptotically non flat solutions due to the global nontrivial topology. Now we obtain the proper celestial coordinates for the asymptotically non-flat solution by abopting \cite{Hioki}
\begin{equation}
\alpha=\lim_{r \to \infty} -r \frac{p^{(\phi)}}{p^{(t)}}
\end{equation}
\begin{equation}
\beta=\lim_{r \to \infty} r \frac{p^{(\theta)}}{p^{(t)}}
\end{equation}
where $(p^{(t)},p^{(r)},p^{(\theta)},p^{(\phi)})$ are the tetrad components of
the photon momentum with respect to locally nonrotating reference frame.  So basically one can define the observer’s sky as the usual cases  in which the observer bases $e^{\mu}_{(\nu)}$  can be expanded as a form in the coordinate bases. In the limit $r \to \infty $ can relate the above coordinates to parameters $\xi$ and $\eta$, which then yield
\begin{eqnarray}
\alpha &=& -\sqrt{ 1-8 \pi \gamma^2}\,\dfrac{\xi}{\sin \theta} \\\label{beta}
\beta &=& \pm \sqrt{ 1-8 \pi \gamma^2} \,\sqrt{\eta + a^2\cos^2\theta -\xi^2\cot^2 \theta}, 
\end{eqnarray}
for the case $\omega=0$ and $\omega=1/3$. And similarly
\begin{eqnarray}
\alpha &=& -\sqrt{ 1-8 \pi \gamma^2-\upsilon}\,\dfrac{\xi}{\sin \theta}\\\label{beta}
\beta &=& \pm \sqrt{ 1-8 \pi \gamma^2-\upsilon} \,\sqrt{\eta + a^2\cos^2\theta -\xi^2\cot^2 \theta} 
\end{eqnarray}
for the case of quintessence i.e., $\omega=-1/3$. We observe that in the dark matter case there is a similar contribution term compared to the global monopole. In the limit $\gamma \to 0$ and $\upsilon \to 0$ we obtain the usual relations for celestial coordinates for the asymptotically flat solution.

\begin{figure}[h!]
 \includegraphics[width=5.cm,height=5.cm]{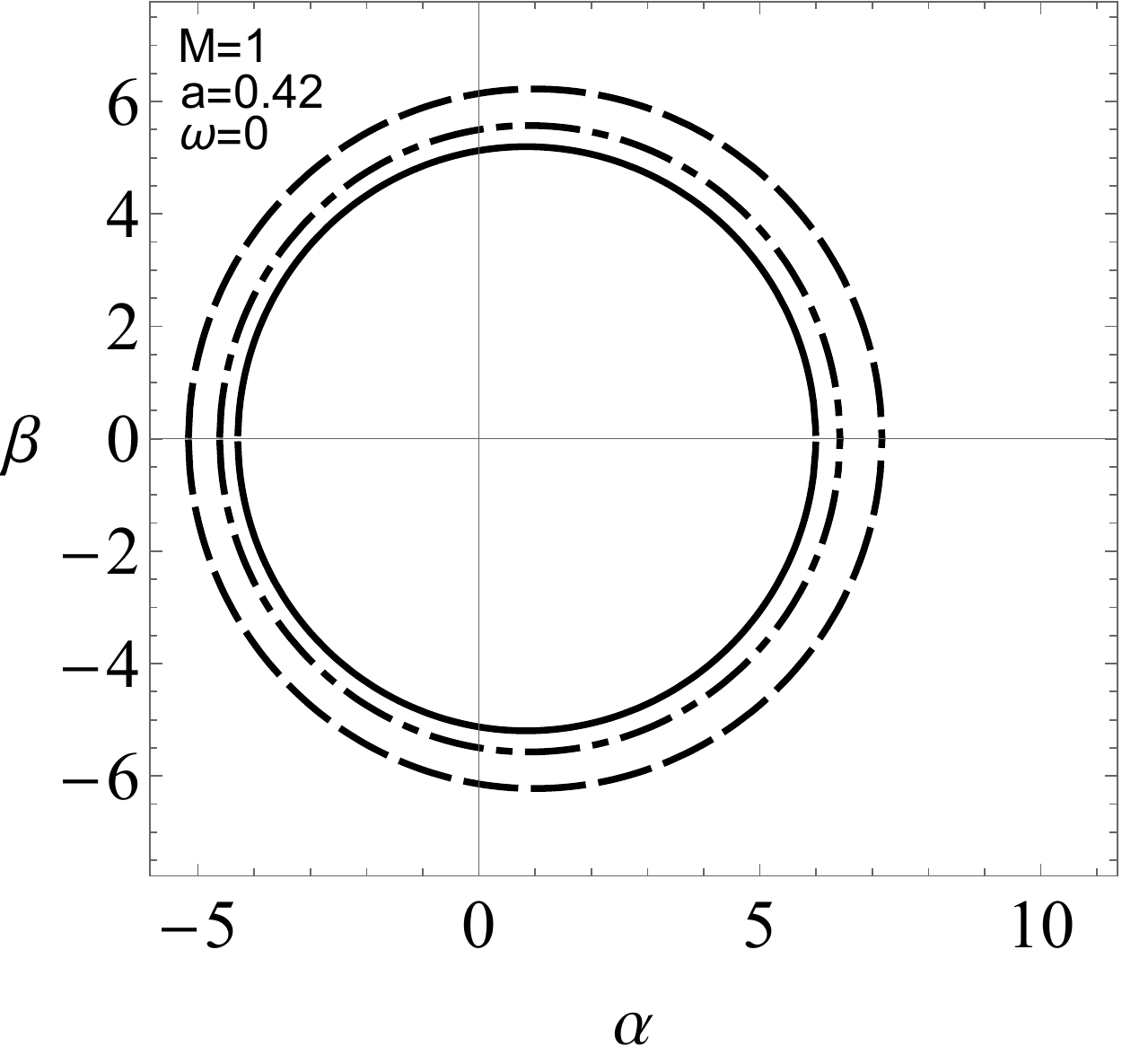}
 \includegraphics[width=5.cm,height=5.cm]{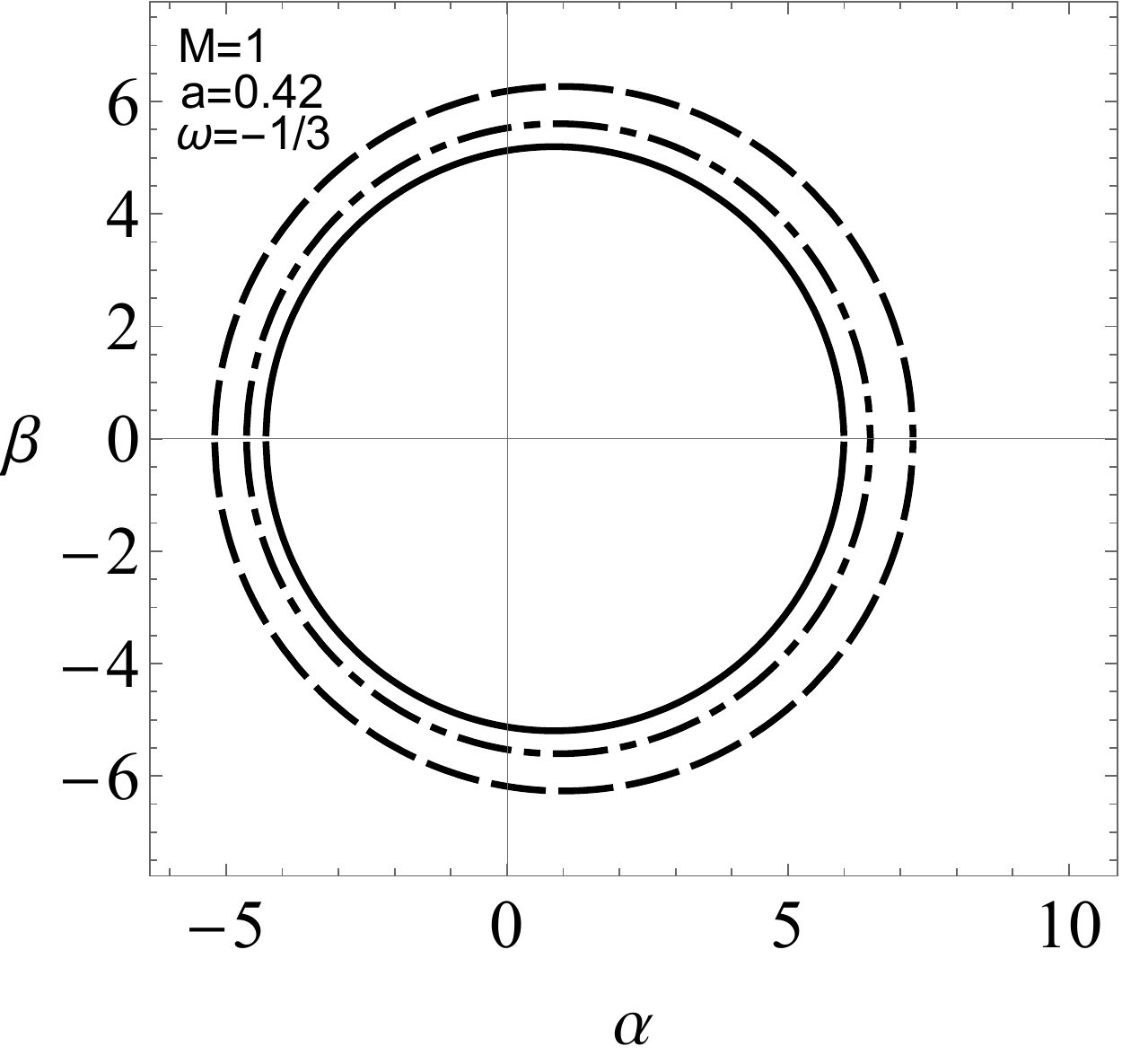}
 \includegraphics[width=5.cm,height=5.cm]{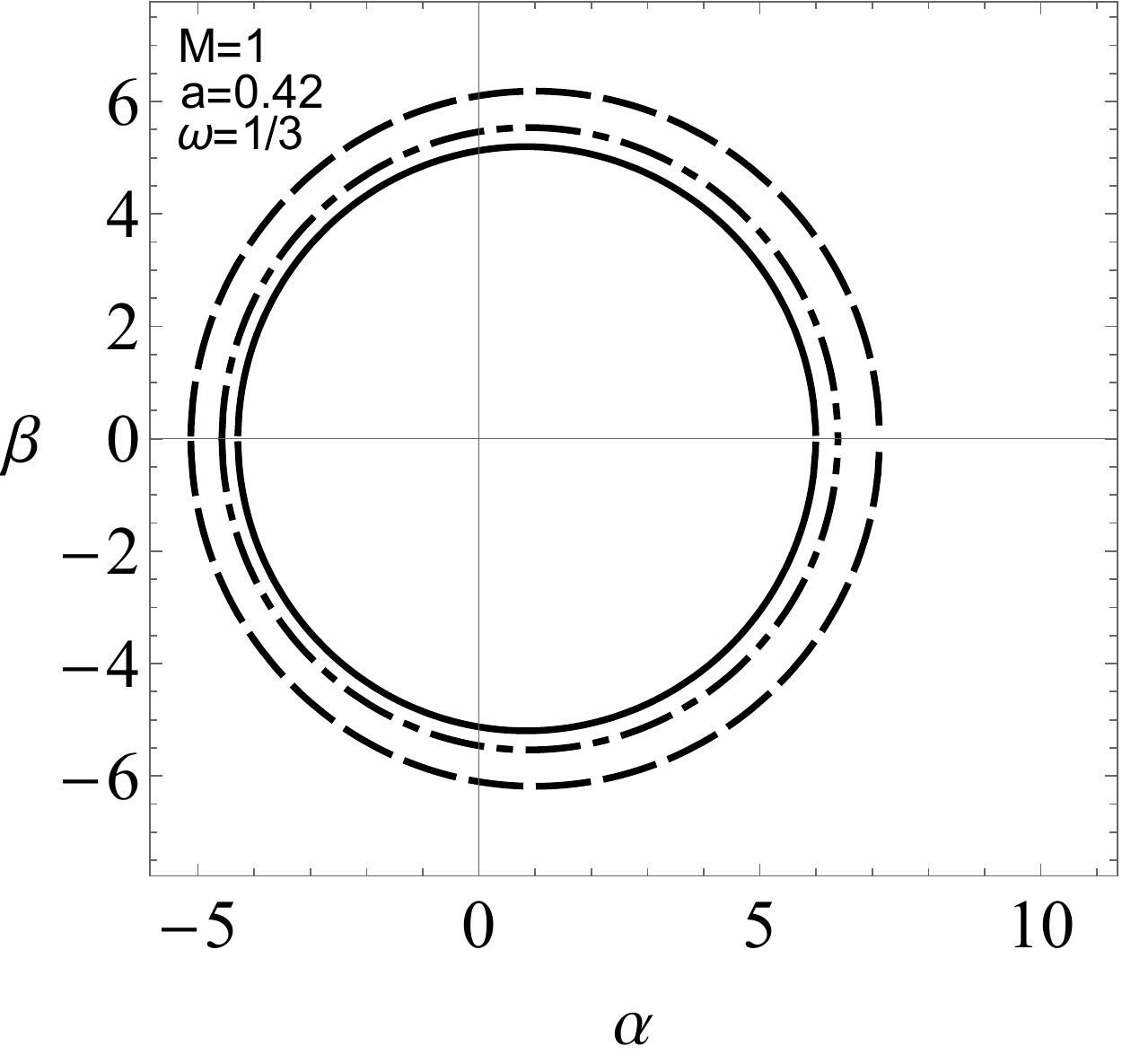}\\
 \includegraphics[width=5.cm,height=5.cm]{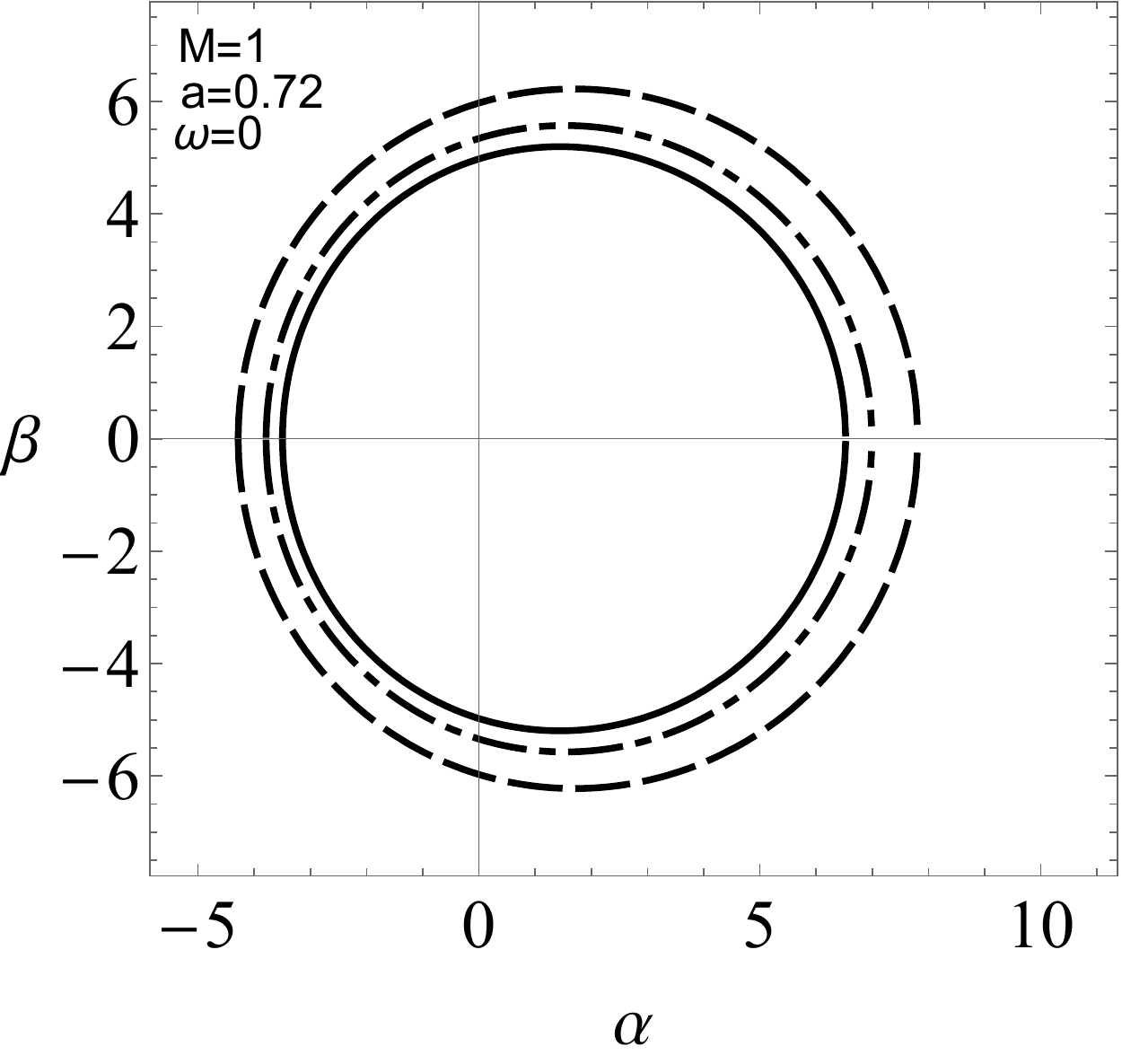}
 \includegraphics[width=5.cm,height=5.cm]{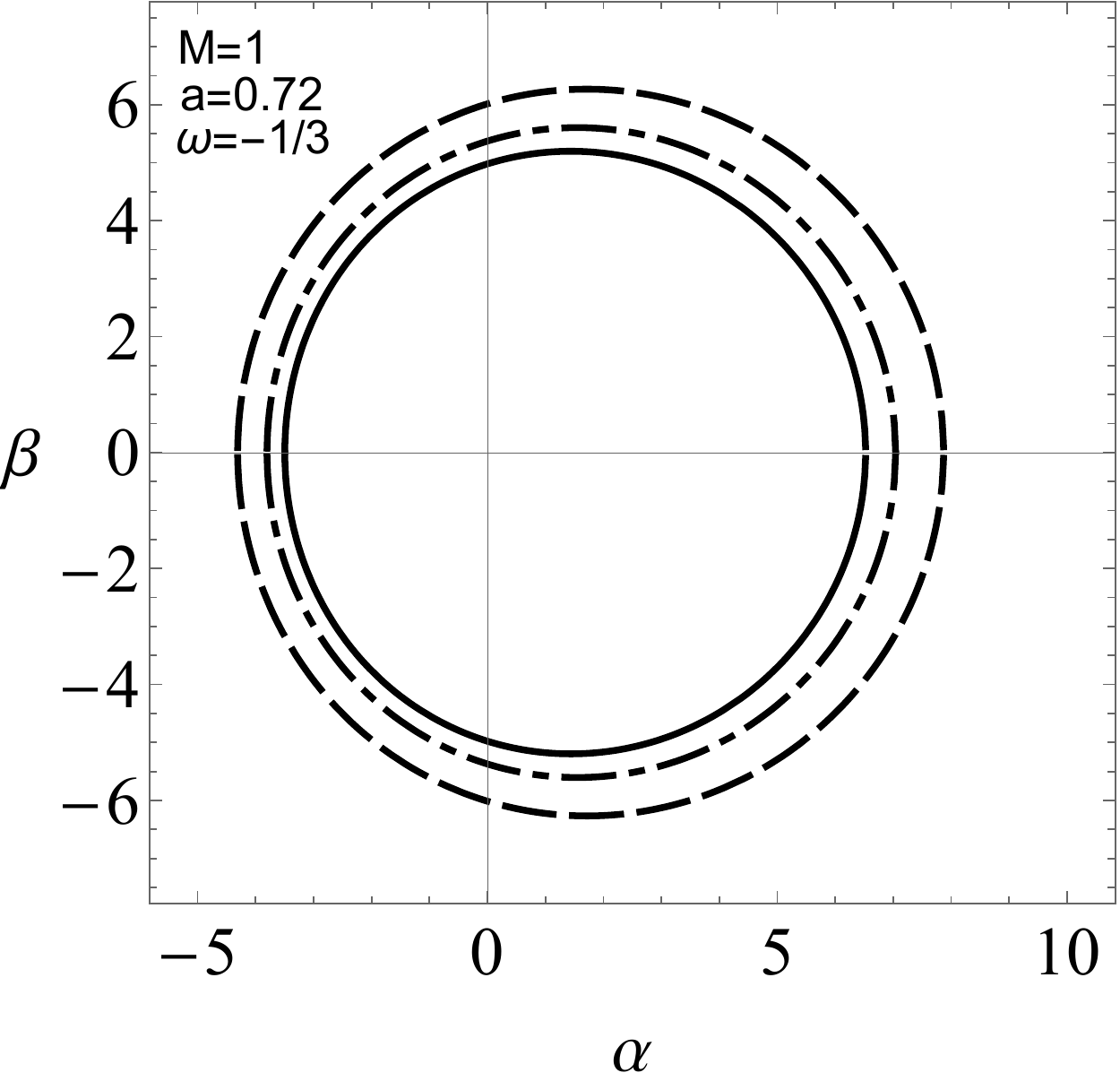}
 \includegraphics[width=5.cm,height=5.cm]{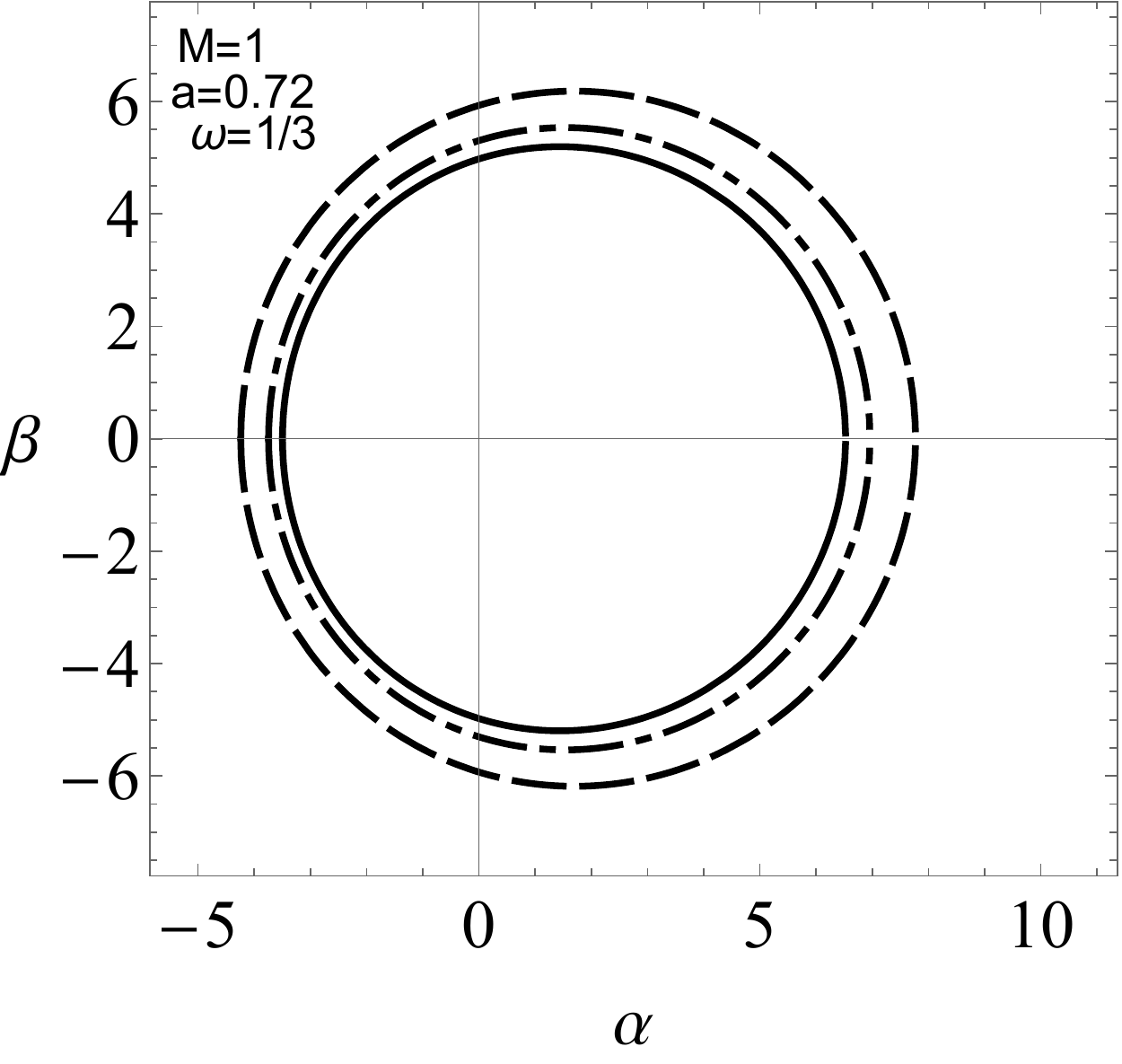}
 \includegraphics[width=5.cm,height=5.cm]{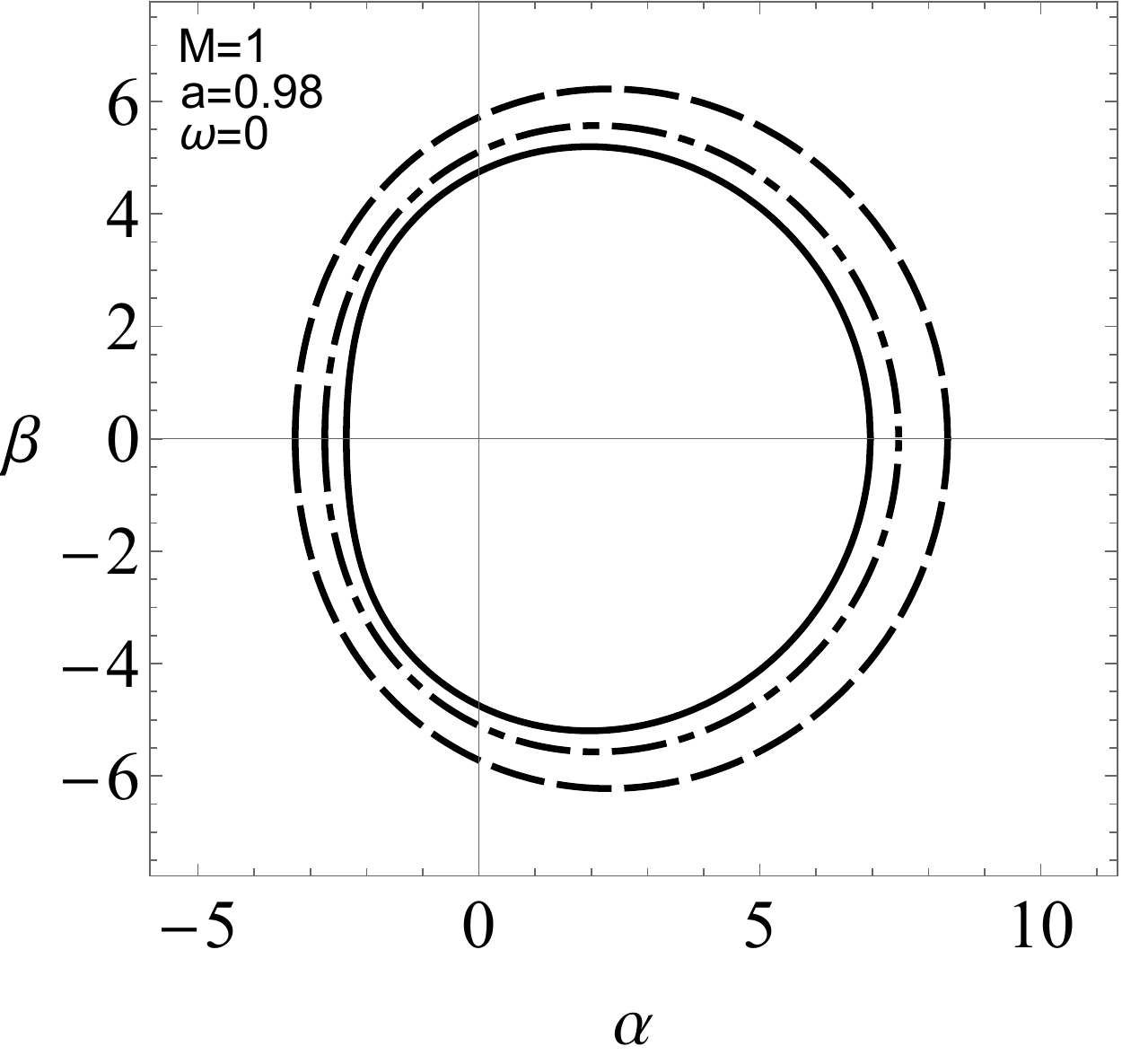}
 \includegraphics[width=5.cm,height=5.cm]{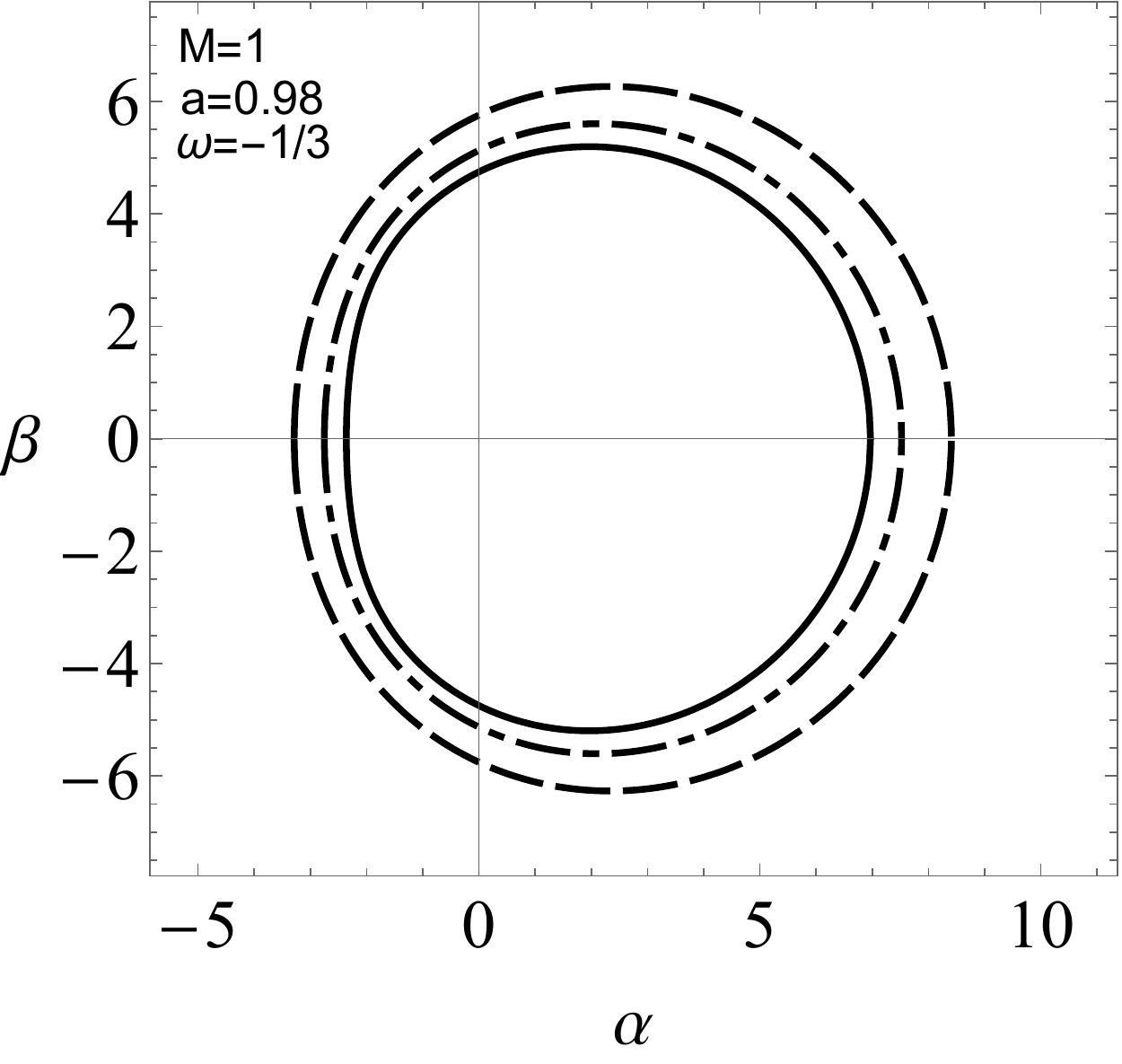}
 \includegraphics[width=5.cm,height=5.cm]{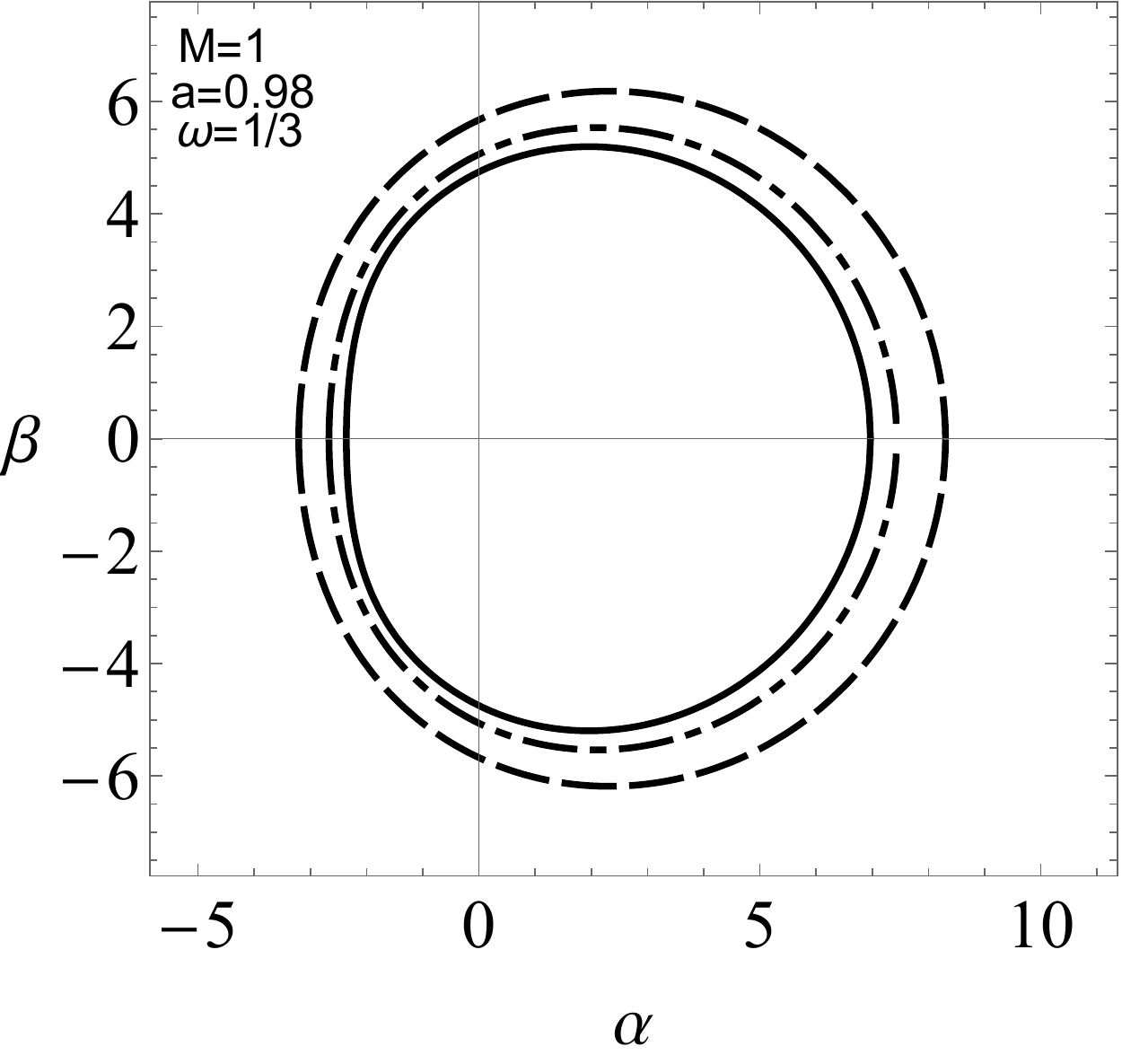}
  \caption{\label{figure1}Variation in shape of a rotating dyonic global monopole surrounded by a perfect fluid. Magnetic and electric charges are kept constant such that $Q_E=10^{-2}=Q_M$. 
   In each graph the Kerr case i.e. $\gamma=0$ and $\upsilon=0$, is represented by solid line, $\gamma=0.05$ by dotdashed and $\gamma=0.08$ by dashed lines.  For dark matter $(\omega=-1/3)$ and dust $(\omega=0)$ case $\upsilon=0.01$,  whereas $\upsilon=-0.01$ in case of radiation $(\omega=1/3) $.
  }
  \end{figure}
  \begin{figure}[h!]
  \includegraphics[width=5.cm,height=5.cm]{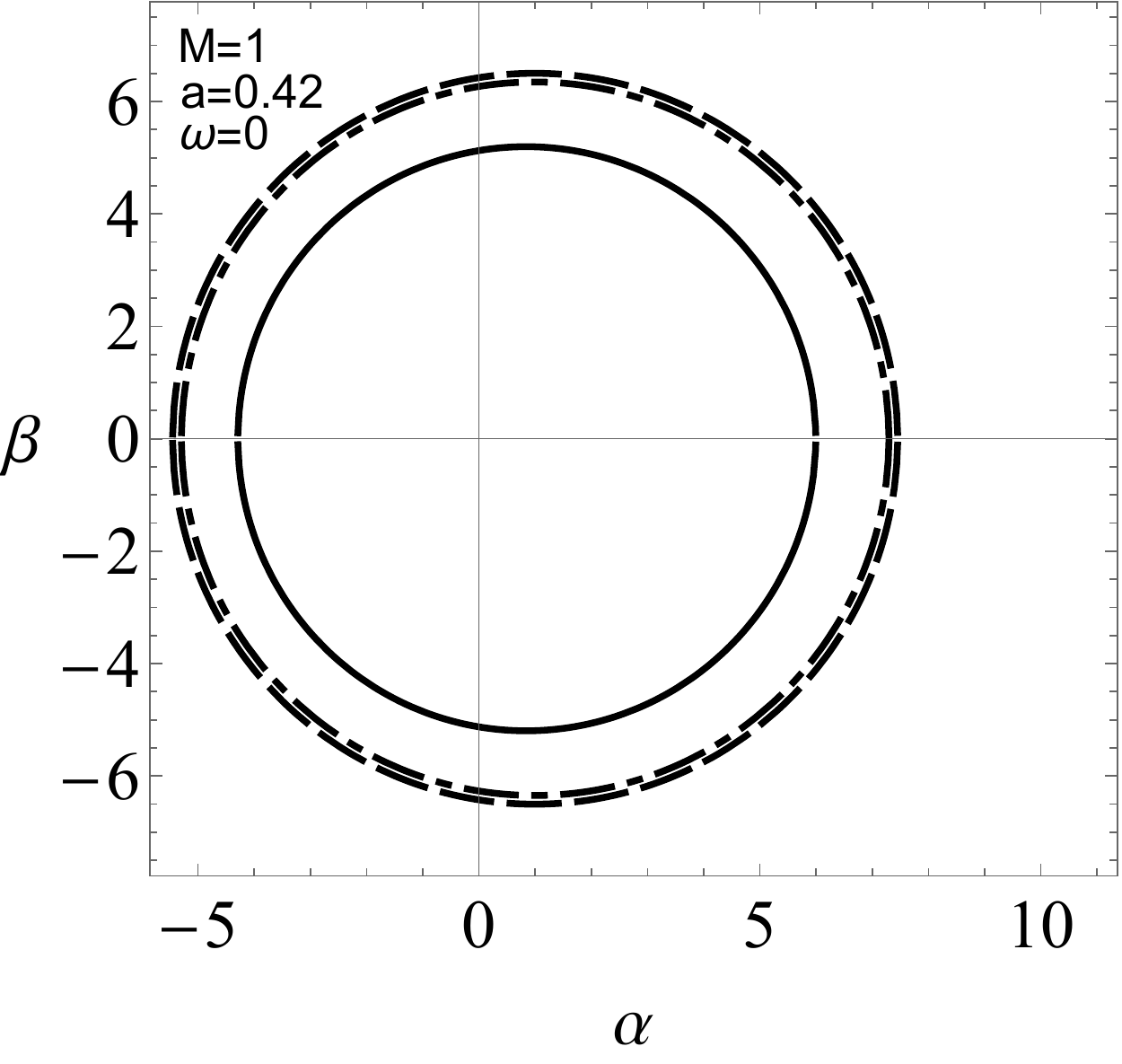}
 \includegraphics[width=5.cm,height=5.cm]{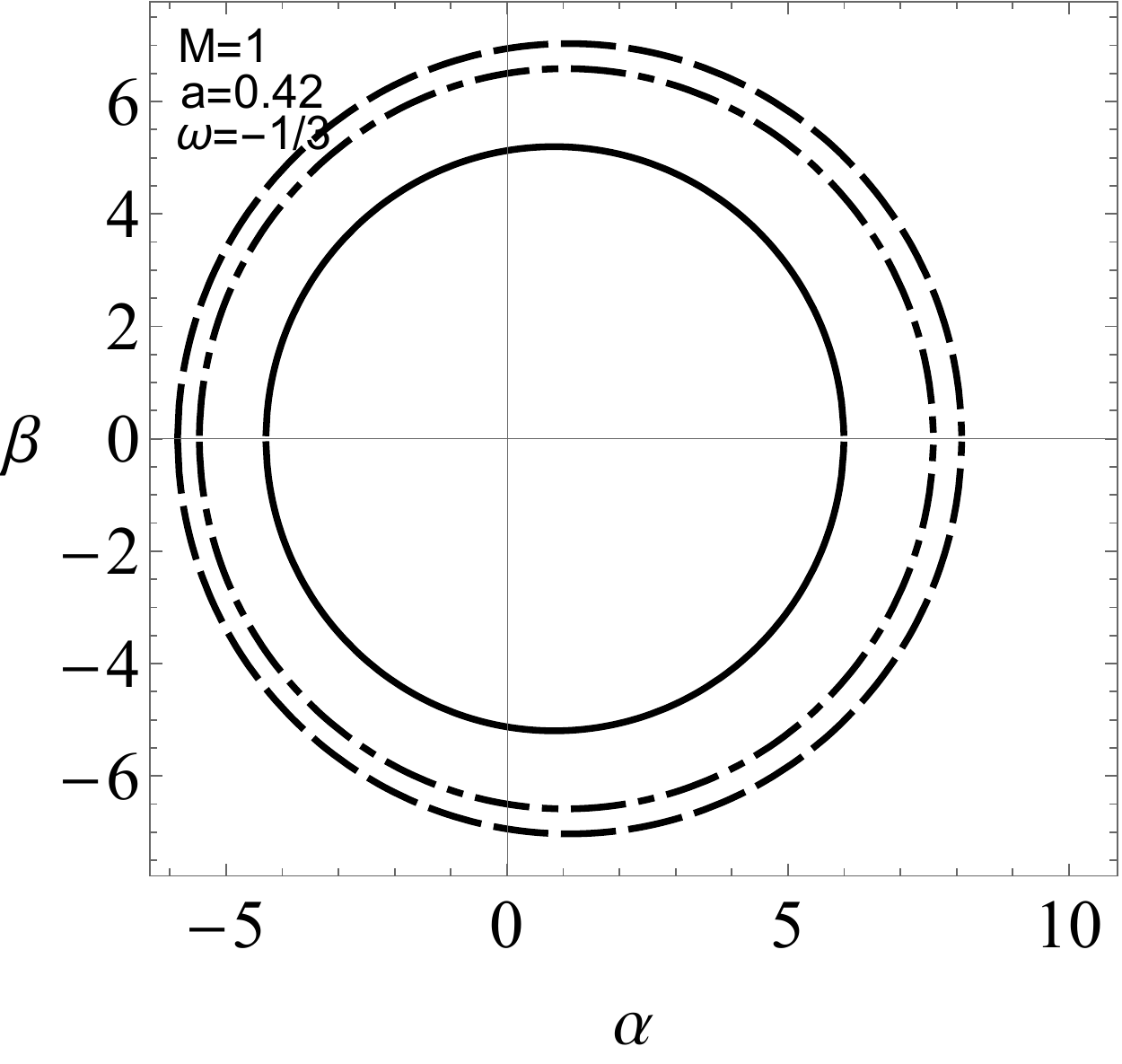}
 \includegraphics[width=5.cm,height=5.cm]{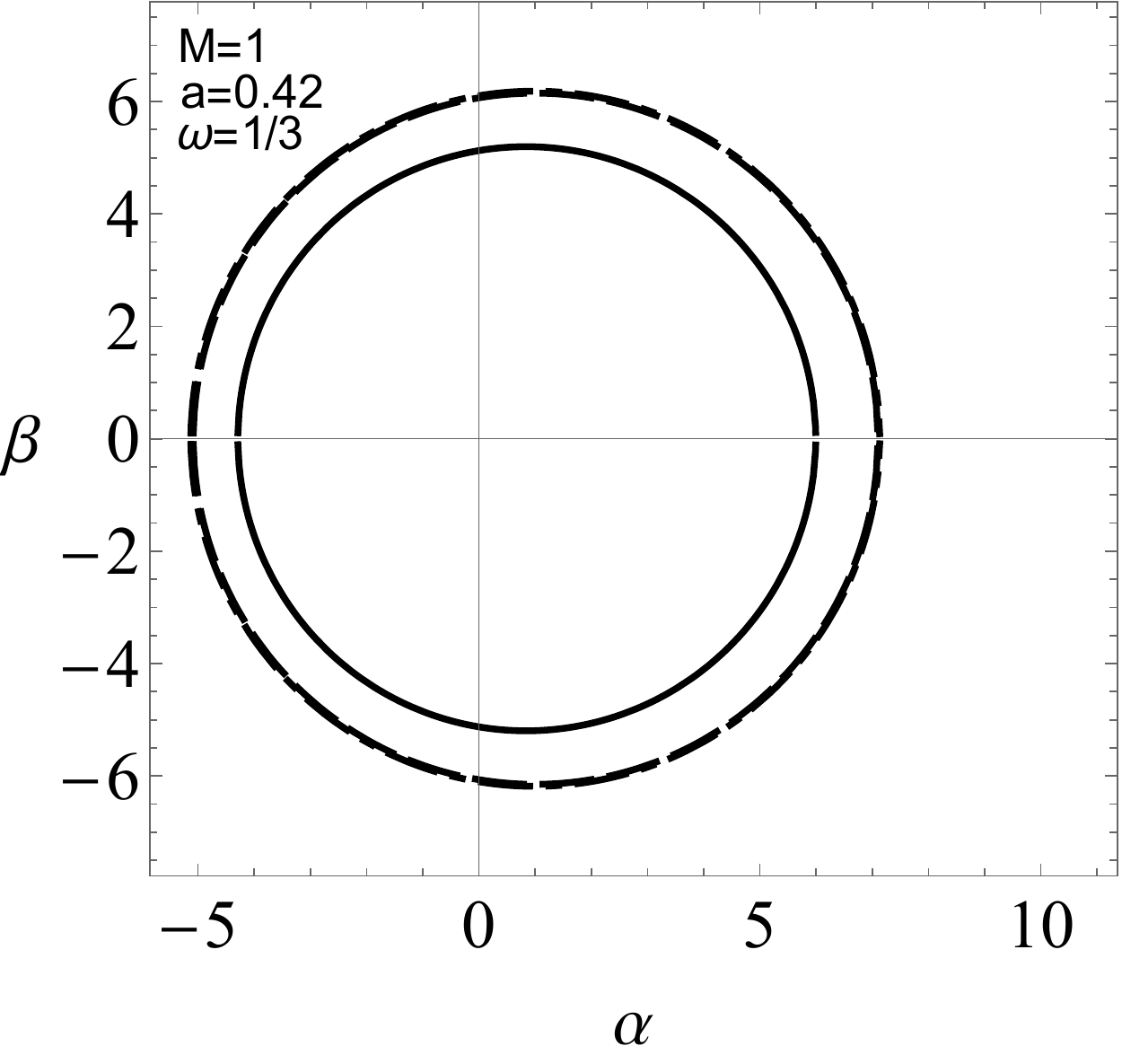}\\
 \includegraphics[width=5.cm,height=5.cm]{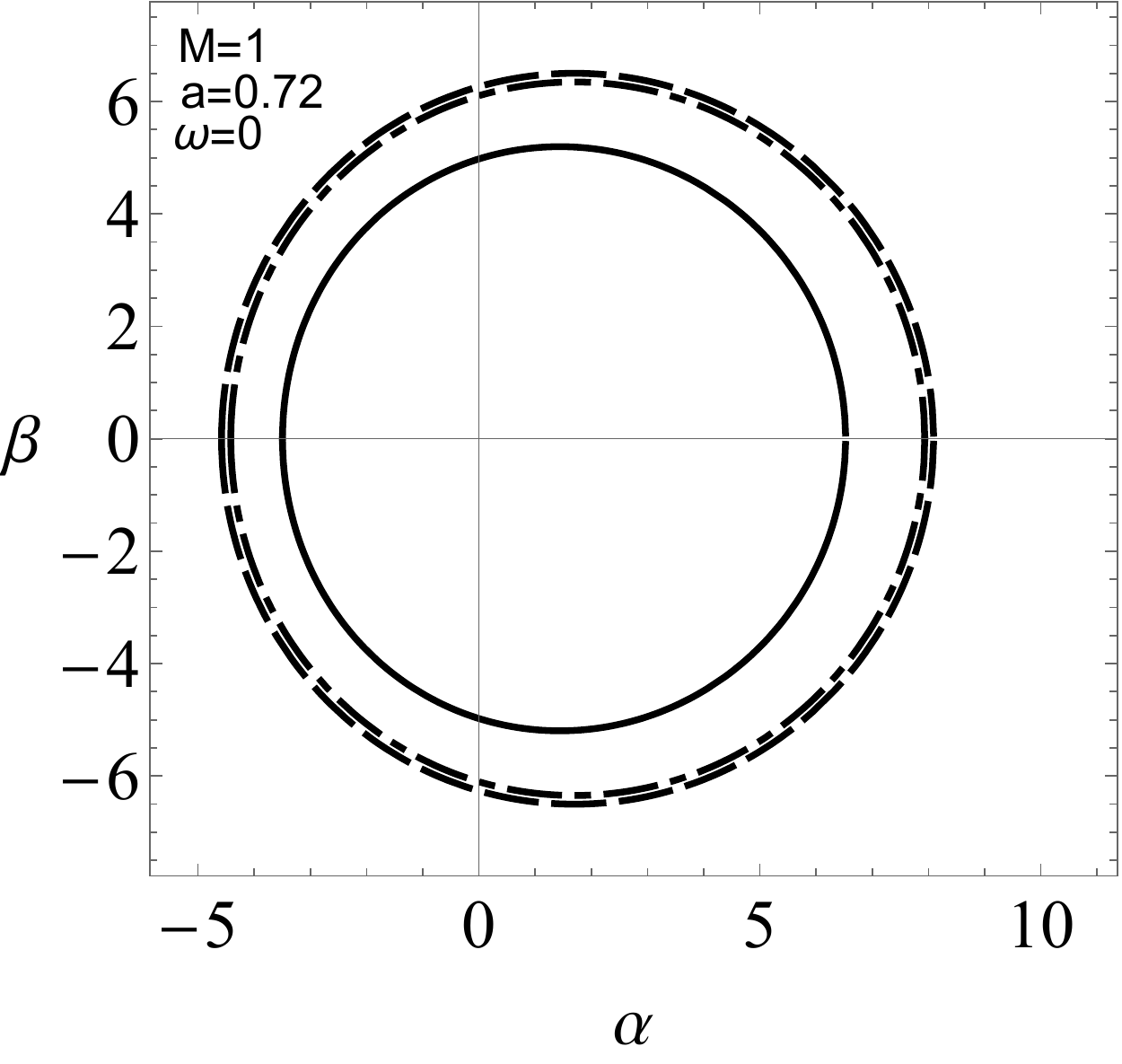}
 \includegraphics[width=5.cm,height=5.cm]{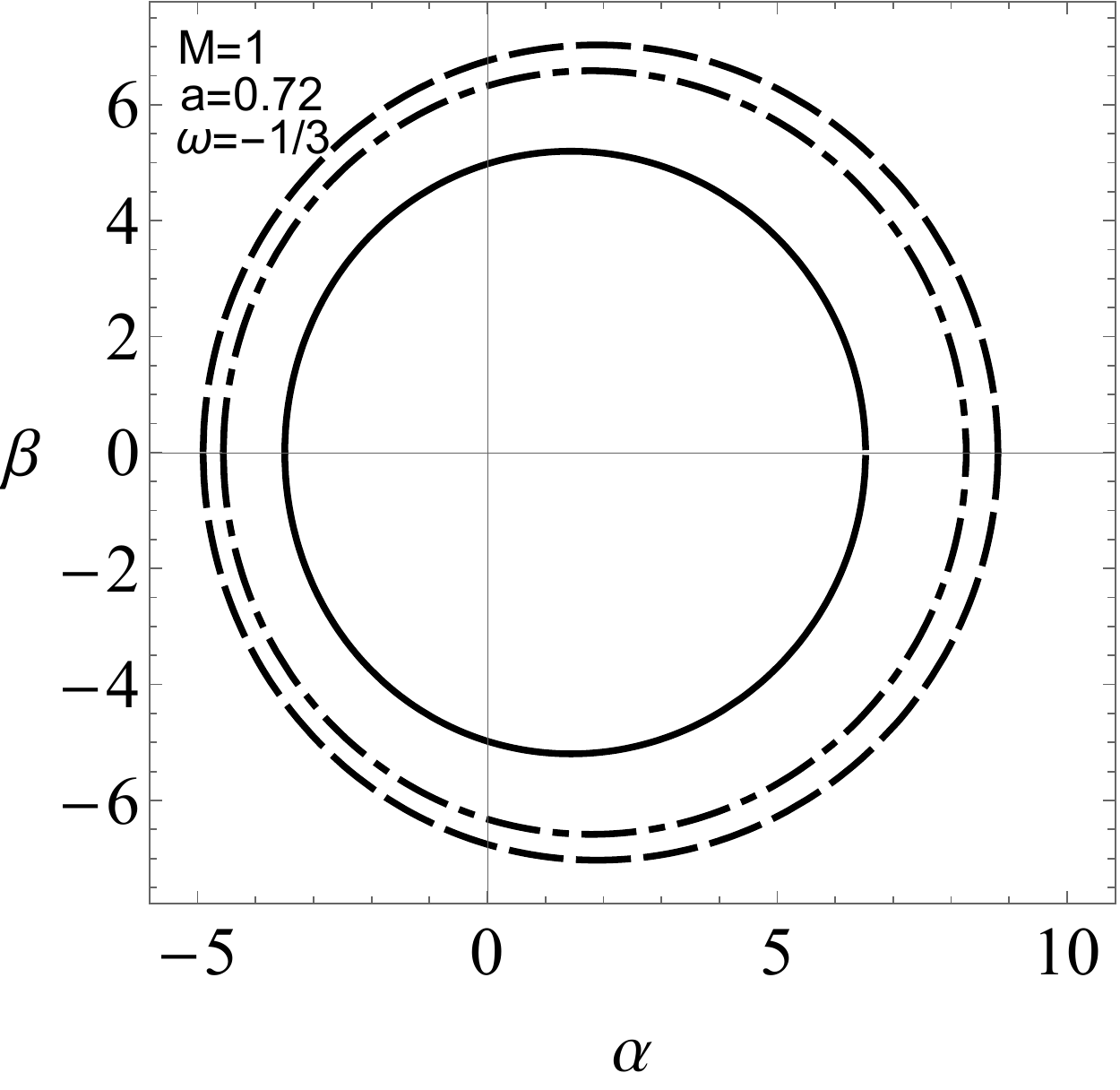}
 \includegraphics[width=5.cm,height=5.cm]{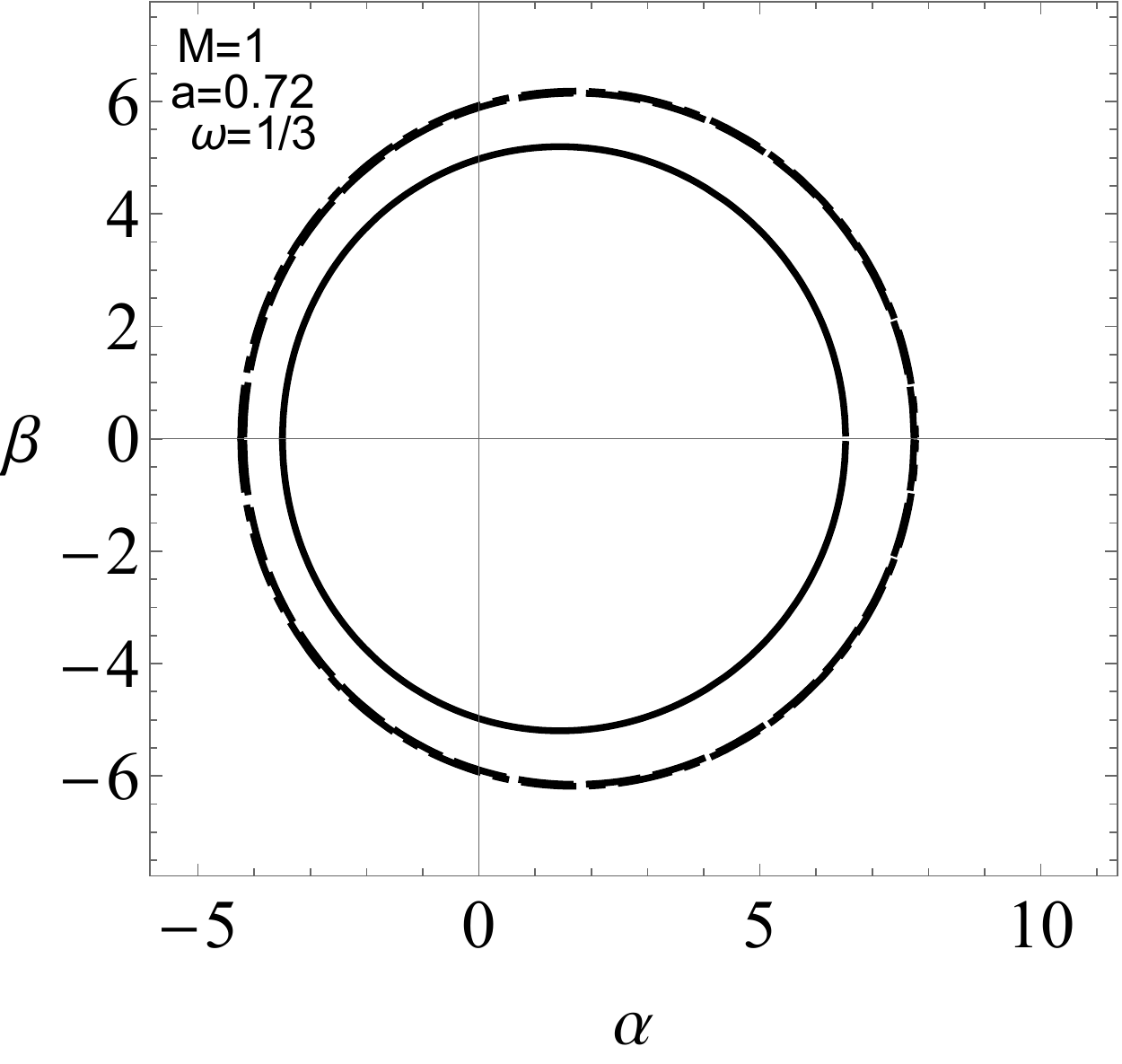}
 \includegraphics[width=5.cm,height=5.cm]{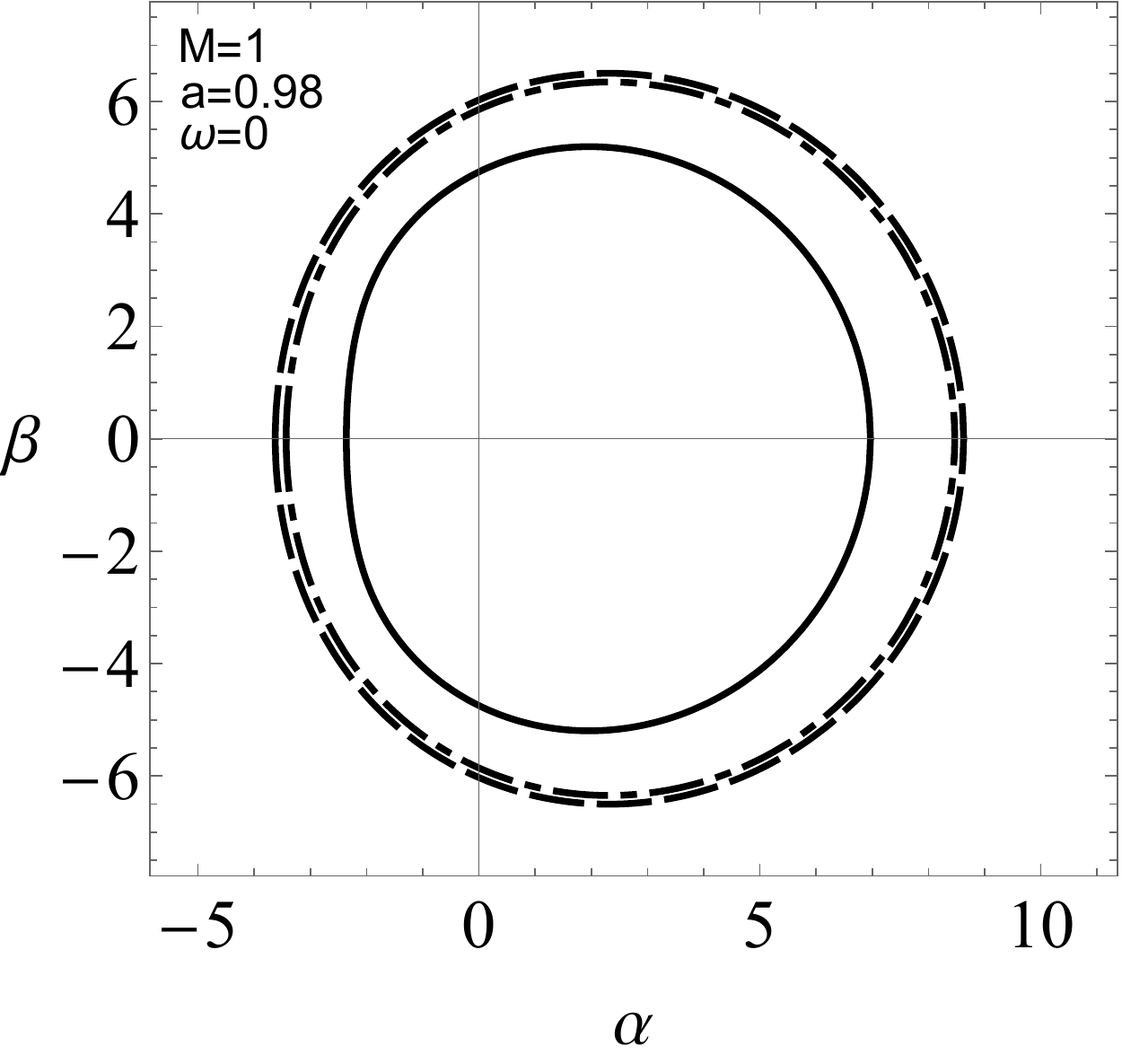}
 \includegraphics[width=5.cm,height=5.cm]{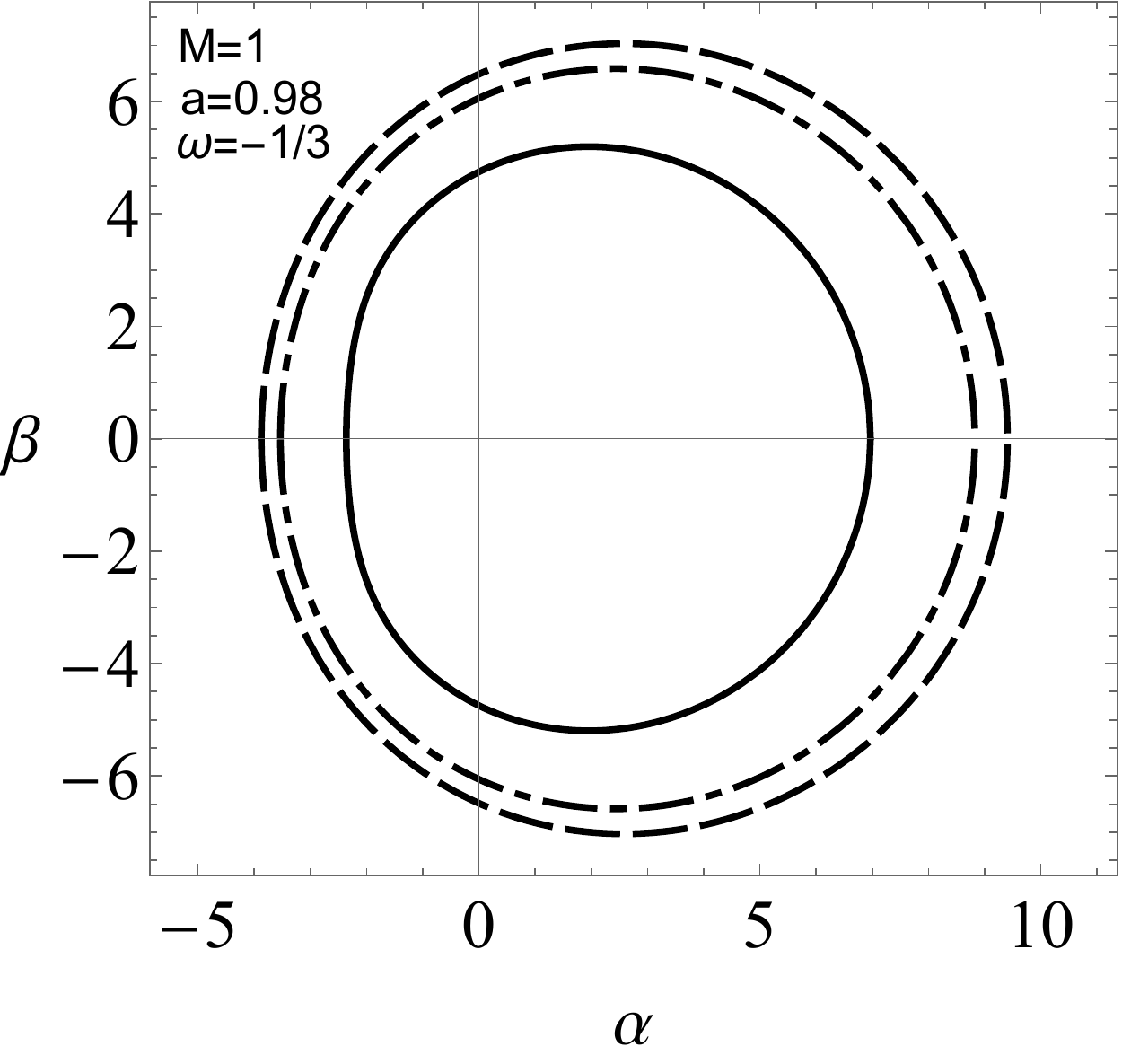}
 \includegraphics[width=5.cm,height=5.cm]{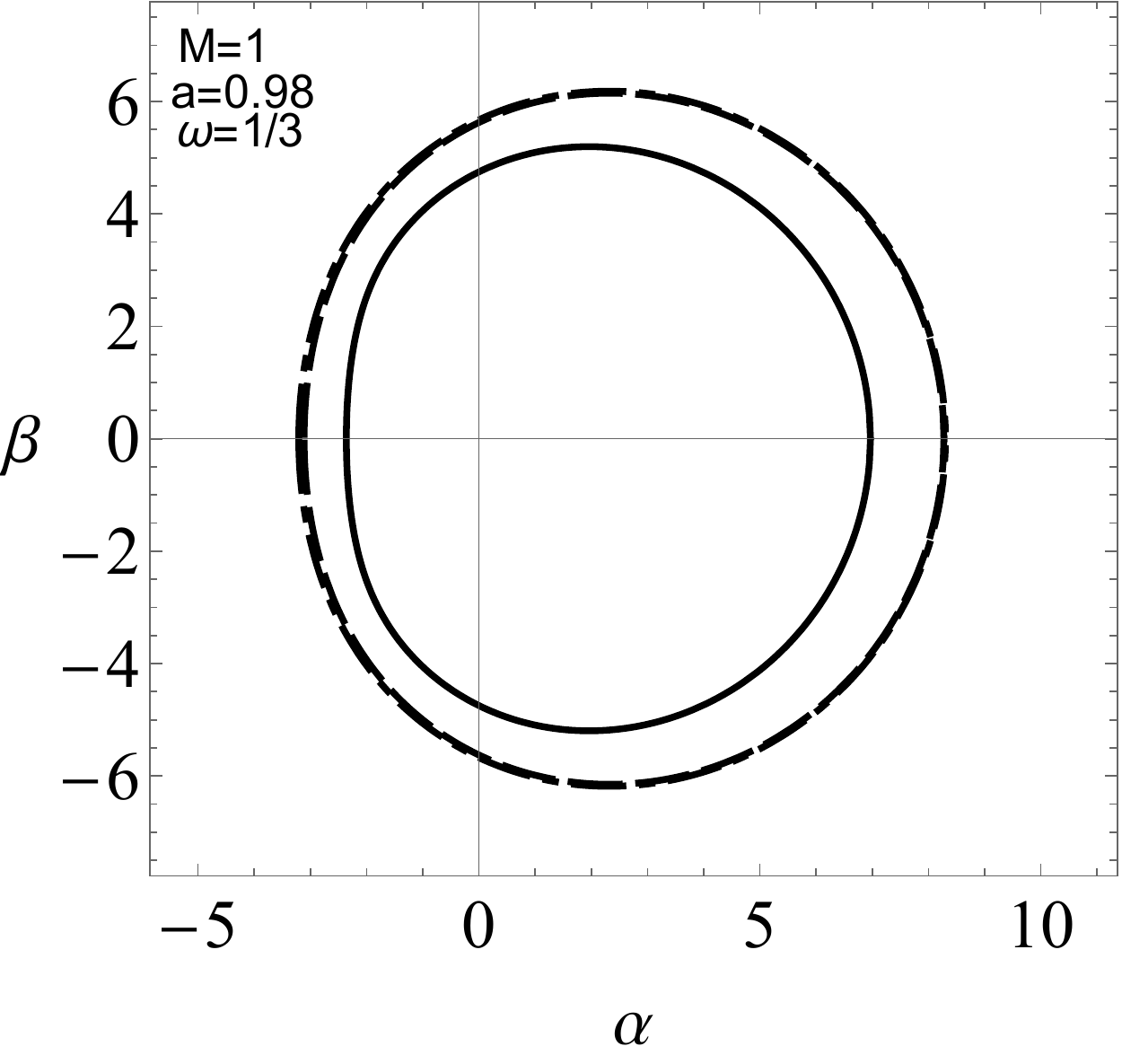}
  \caption{\label{figure2}Variation in shape of a rotating dyonic black hole with global monopole surrounded by a perfect fluid, for different values of perfect fluid parameter $\upsilon$. Magnetic and electric charges along with the global monopole parameter are kept constant such that: $Q_E=10^{-2}=Q_M$ and $\gamma=0.08$. For dark matter and dust case $\upsilon=0$ (Solid), $0.05$ (DotDashed) and $0.1$ (Dashed). In case of radiation $\upsilon=0$ (Solid), $-0.01$ (DotDashed) and $-0.05$ (Dashed).}
  \end{figure}
\begin{figure}[h!]
 \includegraphics[width=7.cm,height=7cm]{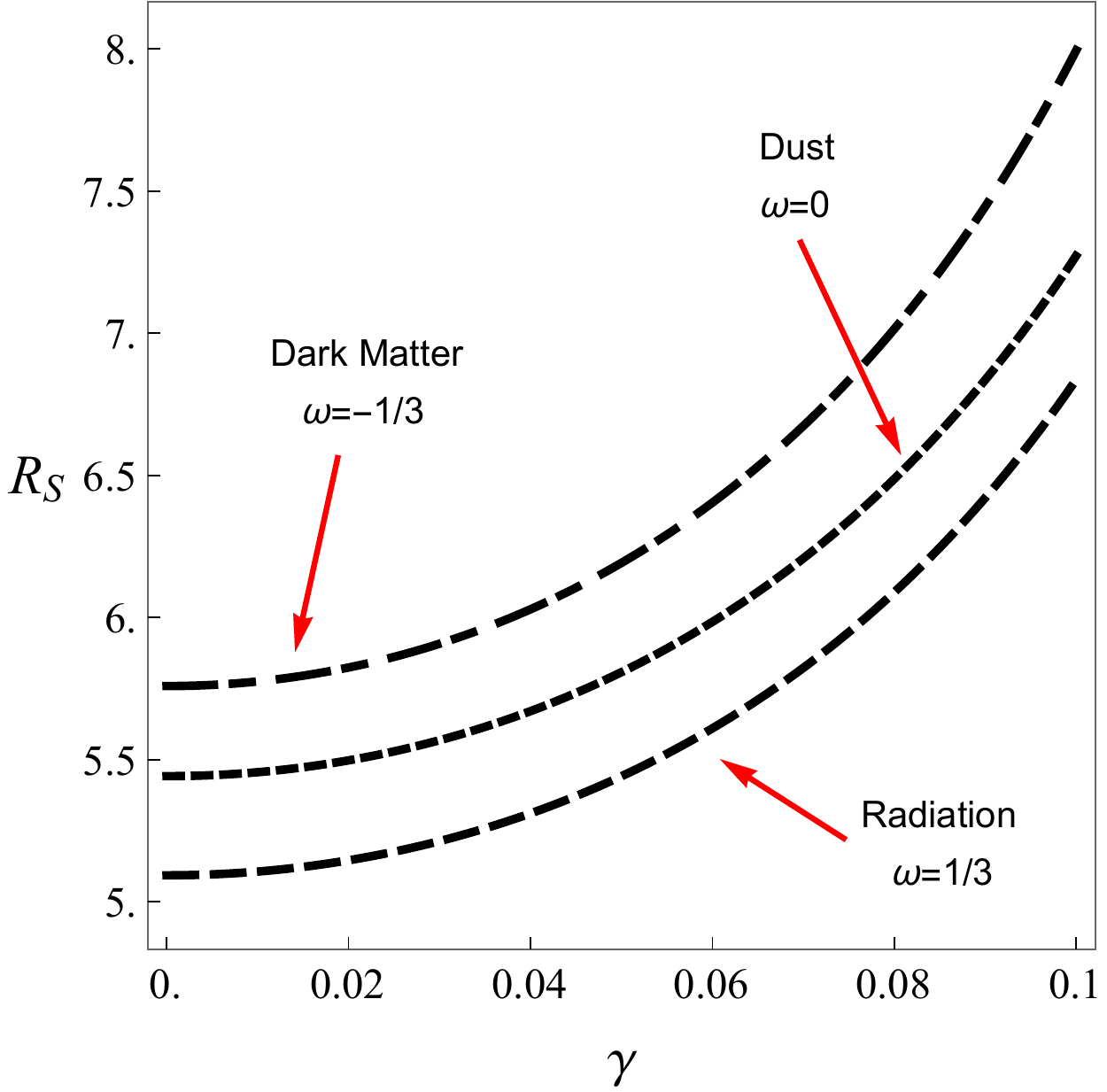}
 \includegraphics[width=7.cm,height=7.cm]{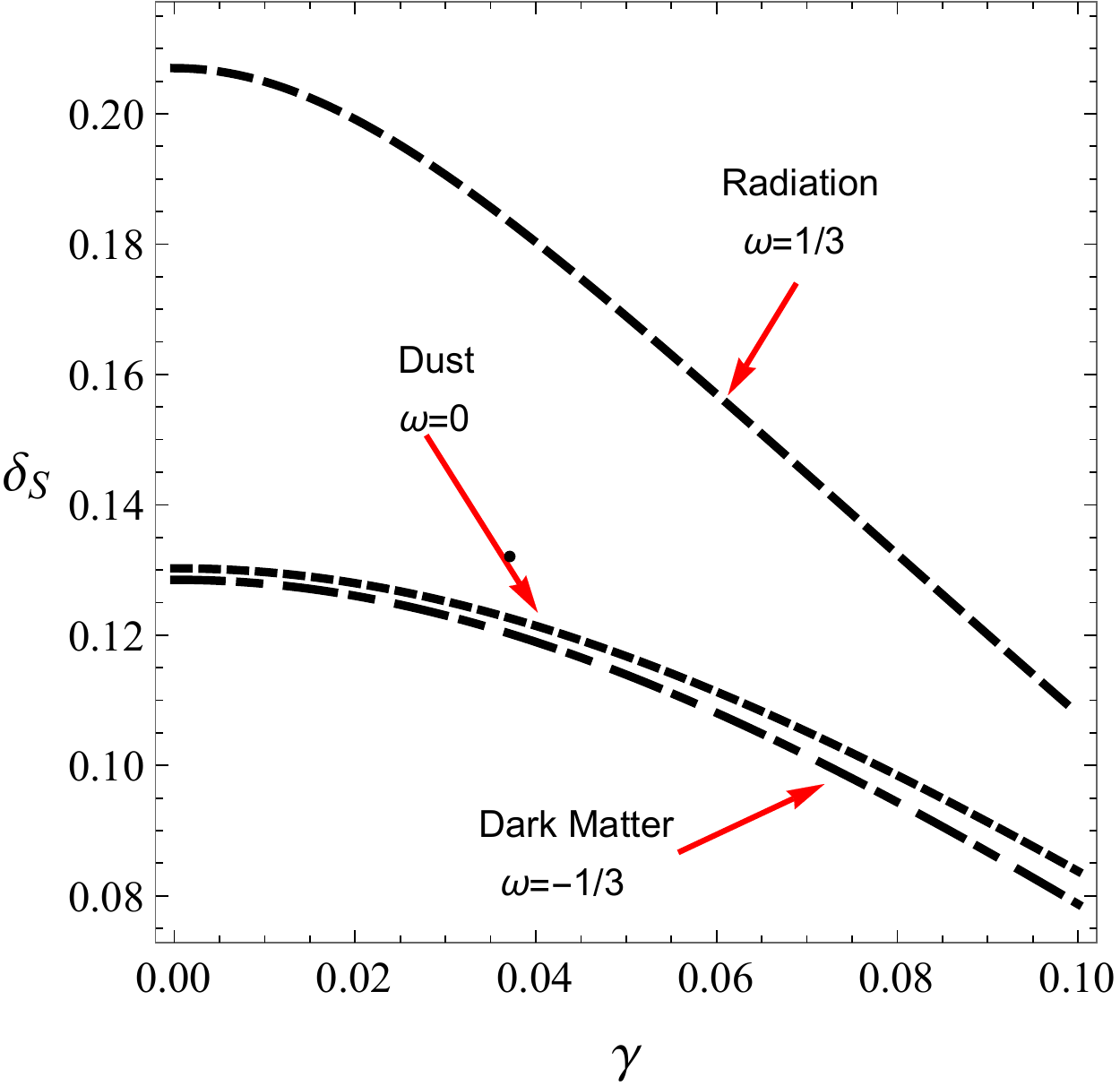}
  \caption{\label{observables}The quantities $R_s$ and $\delta_s$ with respect to parameter $\gamma$}
  \end{figure}
\begin{figure}[h!]
 \includegraphics[width=5cm,height=5cm]{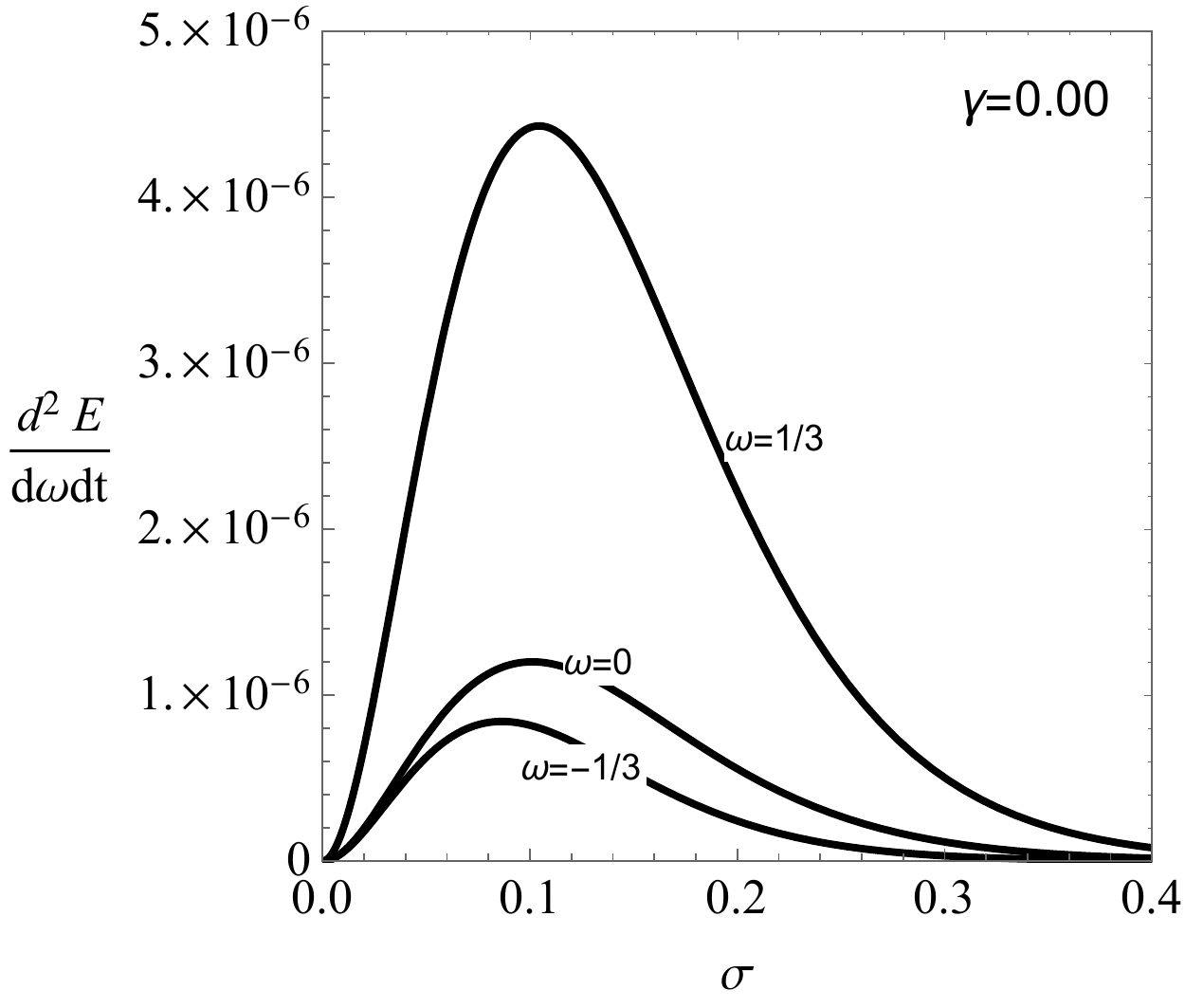}
 \includegraphics[width=5cm,height=5cm]{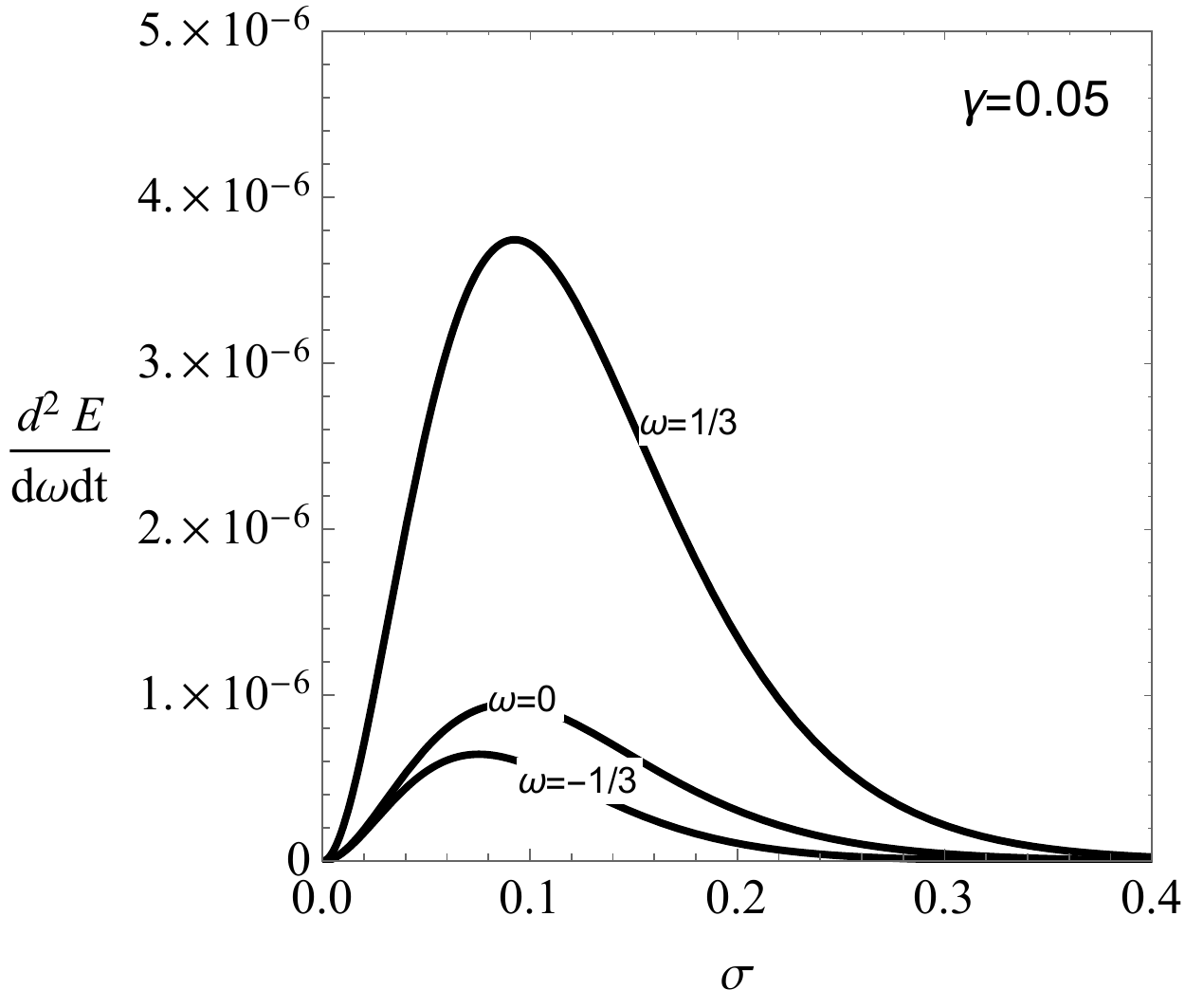}
 \includegraphics[width=5cm,height=5cm]{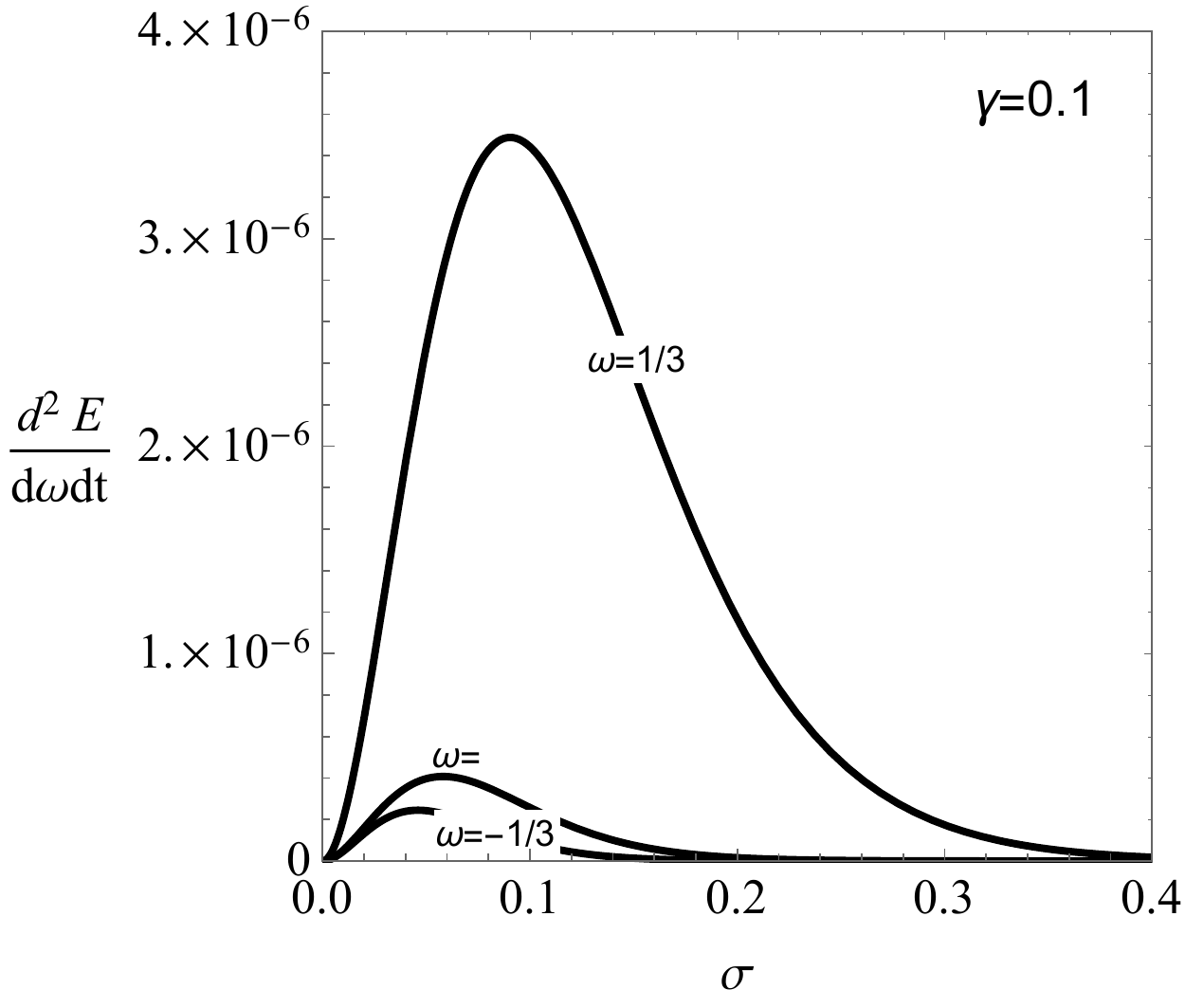}\\
 \includegraphics[width=5cm,height=5cm]{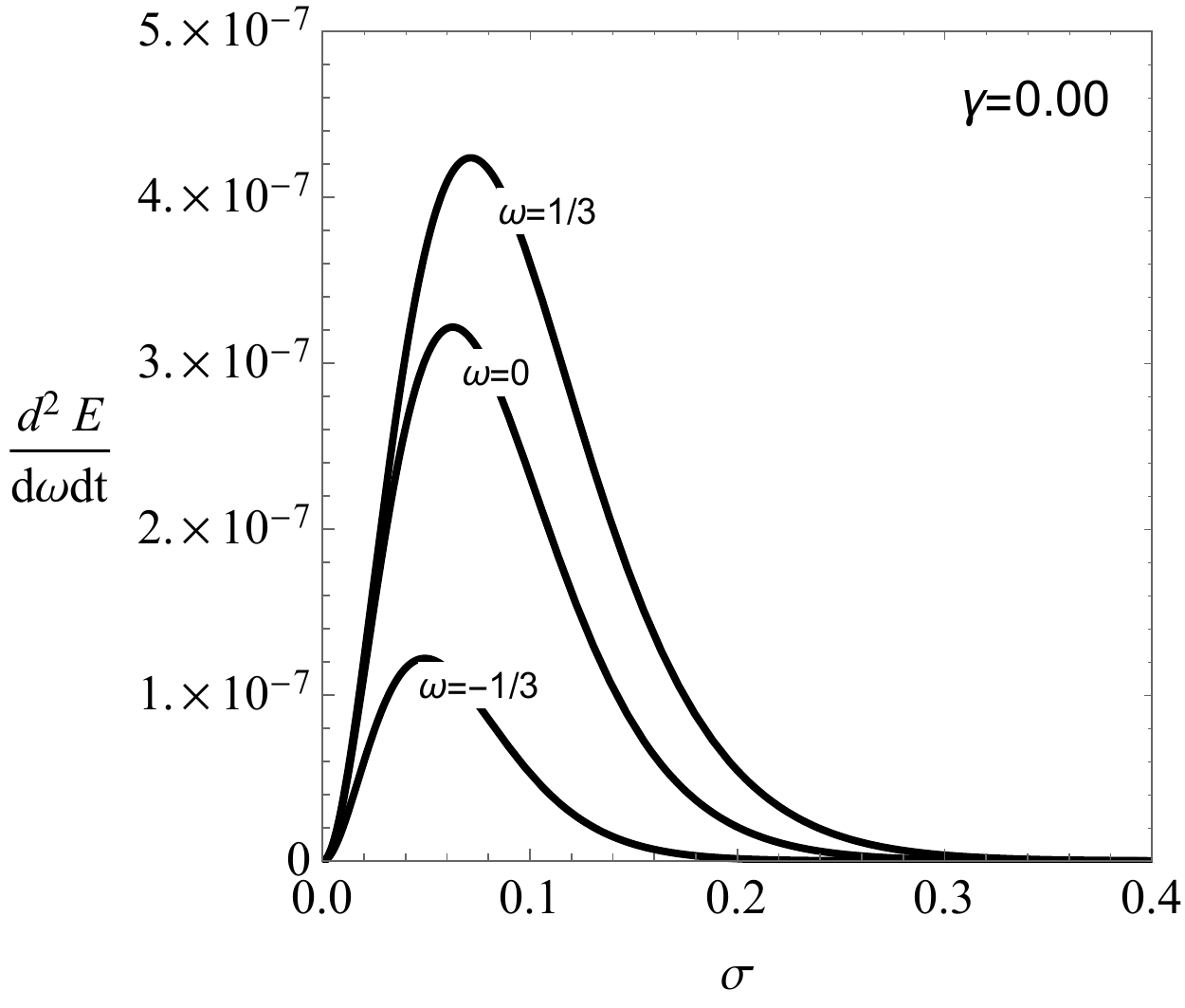}
 \includegraphics[width=5cm,height=5cm]{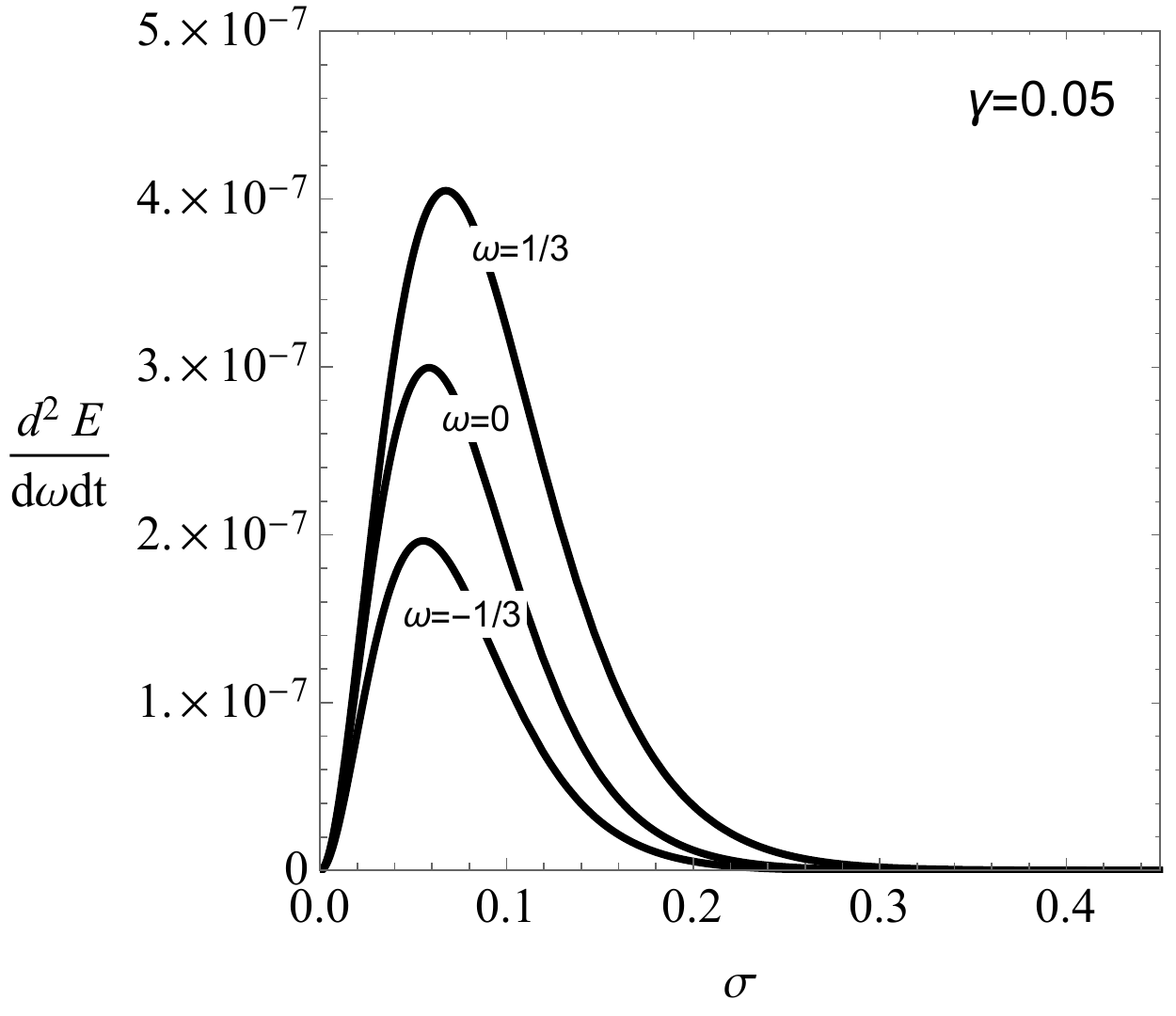}
 \includegraphics[width=5cm,height=5cm]{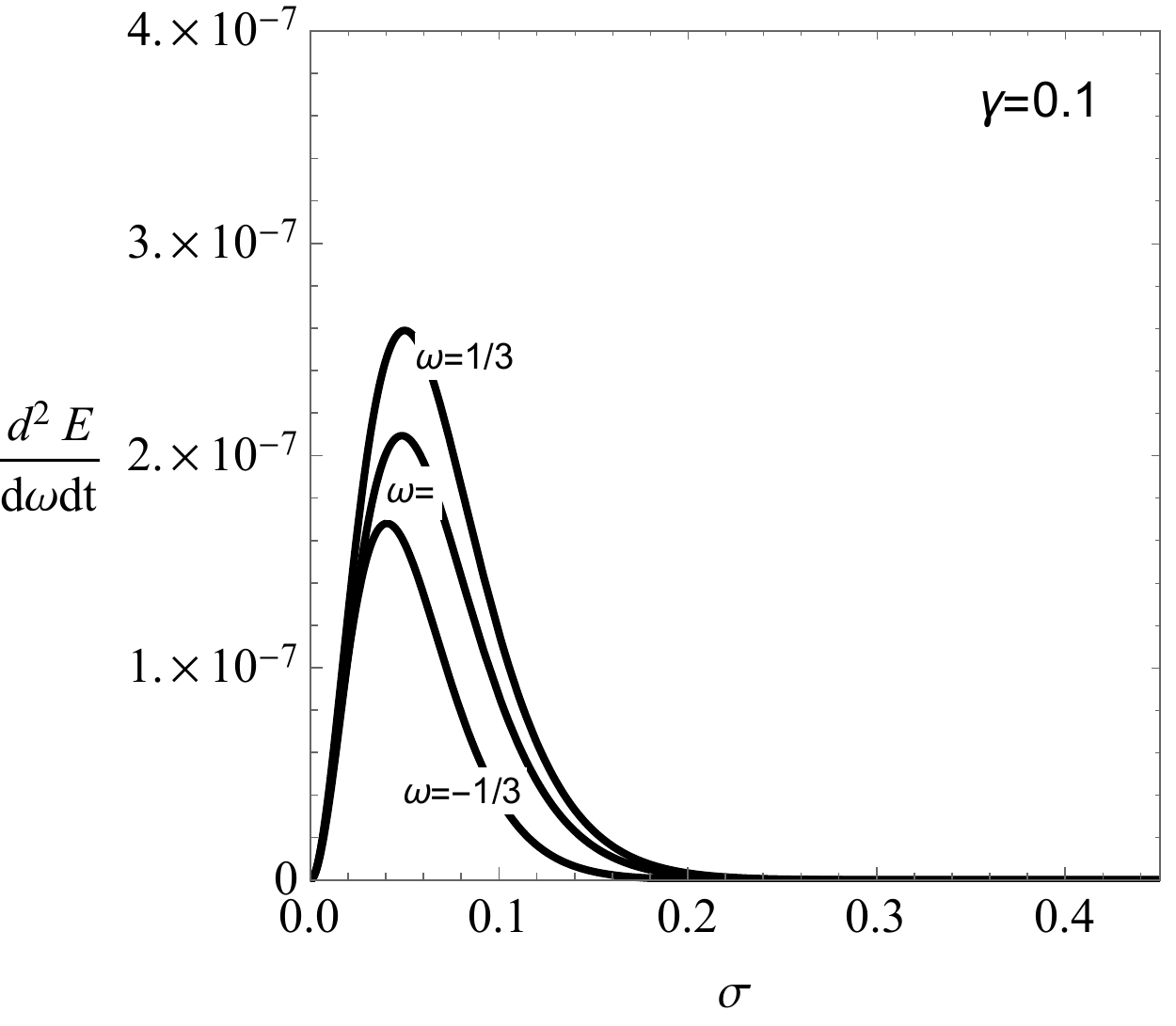}
  \caption{\label{figure4}The figure shows the energy emission rate when $a=0.46$ (upper panel) and $a=0.92$ (lower panel).}
  \end{figure}

We expect that the parameters involve in RDGM in presence of a perfect fluid will effect the shape of its shadow. This can be clearly confirmed through Eq. \ref{beta} as it depends not only on spin parameter $a$ and angular coordinate $\theta_o$ but also on $\gamma$, $\omega$ and perfect fluid parameter $\upsilon$. Later, we will justify our results also through graphical interpretations.\\
As our observer is placed in the equatorial plane ($\theta=\pi/2$), $\alpha$ and $\beta$ reduce to
\begin{eqnarray}
\alpha &=& -\sqrt{ 1-8 \pi \gamma^2}\,\,\,\xi \\
\beta &=& \pm \sqrt{ 1-8 \pi \gamma^2} \,\sqrt{\eta}, 
\end{eqnarray}
for the case $\omega=0$ and $\omega=1/3$. And 
\begin{eqnarray}
\alpha &=& -\sqrt{ 1-8 \pi \gamma^2-\upsilon}\,\,\,\,\xi\\
\beta &=& \pm \sqrt{ 1-8 \pi \gamma^2-\upsilon} \,\sqrt{\eta} 
\end{eqnarray}
for the case $\omega=-1/3$.
Figure (\ref{figure1}) and (\ref{figure2}) show deformation in shapes of the shadow with respect to monopole parameter $\gamma$ and and perfect fluid parameter $\upsilon$, respectively. It is a well known observation now that the rotational effect in a black hole distorts its shape. That being said, we notice in Figure (\ref{figure1}) that for small spin parameter, $a$, the shadow of the black hole maintains a circular shape along with the increase in its size with the inclination of $\gamma$. As for larger spin value, the shadow is clearly distorted and matches with its Kerr counterpart in perfect fluid \cite{Xu} for $\gamma=0$. Figure (\ref{figure2}) shows the effect of parameter $\upsilon$ on the rotating dyonic black hole with a global monopole present in perfect fluid. It is noticed in Figure \ref{figure2} that as perfect fluid parameter, $\upsilon$, increases the size of the shadow also increases. A distortion is noticed in shape of the shadow when the spin parameter $a$ is increased. Also, in case of dark matter and dust, there is significant change in the size of the shadow with respect to $\upsilon$. On the other hand, in case of radiation we do not observe any significant effect of perfect fluid parameter $\upsilon$, in fact the effect is negligibly small.   \\
In \cite{Hioki}, the authors introduce two observables, radius $R_s$ and distortion $\delta_s$, to analyze the size and form of the shadow. The first observable $R_s$ is the approximate radius of the shadow. It is defined by considering a reference circle passing through three points on the boundary of the shadow, such that $(\alpha_\text{\footnotesize{tp}},\beta_\text{\footnotesize{tp}})$ is the top most point on the shadow,  $(\alpha_\text{\footnotesize{bm}},\beta_\text{\footnotesize{bm}})$ is the bottom most point on the shadow and $(\alpha_\text{r},0)$ is the point corresponding to unstable circular orbit seen by an observer on reference frame. Thus
\begin{equation}\label{observable}
R_s=\frac{(\alpha_\text{\footnotesize{tp}}-\alpha_r)^2+\beta_\text{\footnotesize{tp}}^2}{2|\alpha_{\text{\footnotesize{tp}}}-\alpha_r|}.
\end{equation}
The second observable $\delta_s$ is the distortion parameter. Let $D_{CS}$ be the difference between the contour of shadow  and reference circle. Then for the point  $(\tilde{\alpha}_p,0)$ lying on the reference circle and the point $(\alpha_p,0)$ lying on the contour of the shadow, $D_{CS}=|\tilde{\alpha}_p-\alpha_p|$. Thus 
\begin{equation*}
\delta_s=\frac{\tilde{\alpha}_p-\alpha_p}{R_s}.
\end{equation*}
For our case, we consider the points $(\tilde{\alpha}_p,0)$ and $(\alpha_p,0)$ to be on the equatorial plane, opposite to the point $(\alpha_\text{r},0)$. The variations in these observables with respect to monopole parameter $\gamma$ are graphically presented in Figure (\ref{observables}). The dependence of $R_s$ on parameter $\gamma$ is such that as $\gamma$ increases the radius $R_s$ also increases. Thus the size of the shadow increases with increase in monopole parameter $\gamma$. Whereas the distortion $\delta_s$ decreases monotonically with an increase in $\gamma$. This tells us that with respect to circumference of reference circle, the shadow of the rotating black hole is significantly distorted for $\gamma\in[0,0.1]$ but for $\gamma>0$ it may not show any distortion and thus we may obtain a perfect circle.  
As we have considered our observer to be at infinity so in this case the area of the black hole shadow will be approximately equal to high energy absorption cross section as discussed in \cite{shao}. For a spherically symmetric black hole the absorption cross section oscillates around $\Pi_{ilm}$, a limiting constant value. For a black hole shadow with radius $R_s$, we adopt the value of $\Pi_{ilm}$  as calculated by \cite{shao}
\begin{equation*}
  \Pi_{ilm}\approxeq ~ \pi R_s^2.
\end{equation*}
The energy emission rate of the black hole is thus defined by
\begin{equation*}
\frac{d^2E(\sigma)}{d\sigma dt}=2 \pi ^2 \frac{\Pi_{ilm}}{e^{\sigma/T}-1}\sigma^3,
\end{equation*}
where $\sigma$ is the frequency of the photon and $T$ represents the temperature of the black hole at outer horizon i.e. $r_+$, given by
\begin{eqnarray*}
  T(r_+)&=&\lim_{r \to r_+}\frac{\partial_r\sqrt{g_{tt}}}{2 \pi \sqrt{g_{rr}}}\\
  &=&\left(2 a^2 \left(f(r)-1\right)+r(r^2+a^2)f^{'}(r)\right)\frac{r}{4 \pi \left(r^2+a^2\right)^2}
\end{eqnarray*}
For all three cases, radiation, dust and dark matter, the energy emission rate is graphically presented in Figure (\ref{figure4}) where we notice that the energy emission rate decreases with increase in parameter $\gamma$. A slight shift to the lower frequency is also observed while $\gamma$ increases. The spin parameter $a$ also effects the shape of the energy emission rate as an abrupt decrease in energy emission rate is noticed for higher spin value.

Recent studies have pointed out a connection between the topology of the shadow shape by introducing new quantity such as the local curvature radius of the shadow. In our paper have applied the Gauss-Bonnet theorem to the horizon surface area to prove that the topology is a 2 sphere. It will be interesting to see if one can find the shadow radius by means of Gauss-Bonnet theorem applied directly to metric by using a relation between the horizon radius and the photon sphere. We are planning to work on such a project in the near future. 
\section{Conclusion}
In this paper we have used the complex transformations pointed out by Newman and
Janis to obtain a RDGM solution in presence of a perfect fluid matter. Using the Gauss-Bonnet theorem we have shown that the surface topology of a RDGM is indeed a 2-sphere. Furthermore by choosing $\omega=-1/3, 0, 1/3$ we have explored the impact of dark matter, dust, radiation, as well as the global monopole parameter $\gamma$, and perfect fluid parameters $\upsilon$, on the silhouette of black hole. We have found that a rotating dyonic black hole with a global monopole retains a circular shape for small spin parameter. Whereas for high spin like $a=0.98 M$ the shadow of RDGM is distorted. Also as monopole parameter $\gamma$ increases, a slight shift towards the right is also noticed in shape of shadow of black hole under consideration. The two observables, $R_s$ and $\delta_s$, are also being discussed. In the end we analyze energy emission rate of rotating dyonic global monopole surrounded by perfect fluid with respect to parameters.\\

\end{document}